\documentclass[preprint2]{aastex}
\usepackage{subfigure}

\def\meth {CH$_3$OH~}
\def\hho  {H$_2$O~}
\def\hii  {\ion{H}{2}~}
\def\hi {\ion{H}{1}}

\def\d    {\ifmmode {{\rlap{.}}^\circ}\else {${\rlap{.}}^\circ$}\fi}
\def\s    {\ifmmode {{\rlap{.}}^{\rm s}}\else {${\rlap{.}}^s$}\fi}
\def\as   {\ifmmode {{\rlap{.}}{''}}\else {${\rlap{.}}{''}$}\fi}
\def\h    {\ifmmode {^{\rm h}}\else {$^{\rm h}$}\fi}
\def\m    {\ifmmode {^{\rm m}}\else {$^{\rm m}$}\fi}

\def\kms  {km~s$^{-1}$}
\def\masy  {mas~yr$^{-1}$}
\def\etal {et al.\ }
\def\eg   {e.g.,~}
\def\ie   {i.e.,~}

\def\vlsr {\ifmmode {V_{\rm LSR}} \else {$V_{\rm LSR}$}\fi}

\newbox\grsign \setbox\grsign=\hbox{$>$} \newdimen\grdimen \grdimen=\ht\grsign
\newbox\laxbox \newbox\gaxbox
\setbox\gaxbox=\hbox{\raise.5ex\hbox{$>$}\llap
     {\lower.5ex\hbox{$\sim$}}}\ht1=\grdimen\dp1=0pt
\setbox\laxbox=\hbox{\raise.5ex\hbox{$<$}\llap
     {\lower.5ex\hbox{$\sim$}}}\ht2=\grdimen\dp2=0pt

\def\lax{\mathrel{\copy\laxbox}}
\slugcomment{2014 July 21}
\shorttitle{Parallaxes of Star Forming Regions in the Scutum Arm} 
\shortauthors{Sato \etal}

\begin{document}

\title{Trigonometric Parallaxes of Star Forming Regions in the Scutum Spiral Arm}

\author{M. Sato\altaffilmark{1}, 
 Y. W. Wu\altaffilmark{1,2,3}, 
 K. Immer\altaffilmark{1},
 B. Zhang\altaffilmark{1}, 
 A. Sanna\altaffilmark{1},
 M. J. Reid\altaffilmark{4}, 
 T. M. Dame\altaffilmark{4},
 A. Brunthaler\altaffilmark{1},
 and K. M. Menten\altaffilmark{1}} 
\altaffiltext{1}{Max-Planck-Institut f\"ur Radioastronomie,
 Auf dem H\"ugel 69, 53121 Bonn, Germany}
\altaffiltext{2}{Purple Mountain Observatory, Chinese Academy of Sciences,
 Nanjing 210008, China} 
\altaffiltext{3}{Graduate University  of Chinese Academy of Sciences, Beijing
100049, China} 
\altaffiltext{4}{Harvard-Smithsonian Center for Astrophysics,
 60 Garden Street, Cambridge, MA 02138, USA}

\begin{abstract}
We report measurements of trigonometric parallaxes for six high-mass star-forming 
 regions in the Scutum spiral arm of the Milky Way as part of the BeSSeL Survey.
Combining our measurements with 10 previous measurements from the BeSSeL Survey
 yields a total sample of 16 sources in the Scutum arm with trigonometric parallaxes
 in the Galactic longitude range from  5$^\circ$ to 32$^\circ$.
{ Assuming a logarithmic spiral model}, we estimate a pitch angle of { $19\d8\pm 3\d1$ } 
 for the Scutum arm, which is larger than pitch angles reported for other spiral arms.
{ The high pitch angle of the arm may be due to the arm's proximity to the Galactic bar.}
The Scutum arm sources show an average peculiar motion of { $4$~\kms\ } slower than the 
Galactic rotation and $8$~\kms\ toward the Galactic center.
While the direction of this non-circular motion has the same sign as determined for sources 
 in other spiral arms,
 { the motion toward the Galactic center is greater for the Scutum arm sources.}

\end{abstract}

 
\section{INTRODUCTION} 

Our Galaxy's spiral structure remains poorly understood due to the lack of 
 accurate distances to massive star-forming regions that trace spiral arms.
In the past, Galactic models have been mostly based on kinematic distances,
 which unfortunately can have large uncertainties due to deviations
 of radial velocities from the assumed circular motions.
However, owing to the high astrometric accuracy achieved by
 Very Long Baseline Interferometry (VLBI), in particular phase-referencing observations,
 it has become feasible to accurately measure
 trigonometric parallaxes of maser sources in massive star-forming regions across the Milky Way.
Since parallax measurements also yield proper motions, when combined with radial velocities 
 and sky positions, one obtains the full three-dimensional spatial location and kinematic 
 properties of the sources.  These allow for a deep understanding of Galactic structure and dynamics.

Recently, many impressive results for Galactic astrometry with VLBI astronomy have been reported
 (\eg \citealt{Reid:09VI, Honma:13}), using the National Radio Astronomy
 Observatory's\footnote[6]{The National Radio Astronomy Observatory (NRAO)
 is a facility of the National Science Foundation operated under
 cooperative agreement by Associated Universities, Inc.}
 Very Long Baseline Array (VLBA), the Japanese VLBI Exploration of Radio Astrometry (VERA) array
 and the European VLBI Network (EVN).  
We are conducting the Bar and Spiral Structure Legacy (BeSSeL)
 Survey\footnote[7]{http://bessel.vlbi-astrometry.org}, an NRAO Key Science project started in 2010
 \citep{Brunthaler:11}.   
The goal of the BeSSeL Survey is to measure accurate distances and proper motions of hundreds
 of high mass star forming regions across the Milky Way.
The results reported here provide locations and motions of six high-mass star-forming regions (HMSFRs)
 in the Scutum spiral arm (also called the Scutum-Centaurus or Scutum-Crux arm), which is the second
 nearest spiral arm inside the Sun's orbit around the Galactic center.

The Scutum arm has been identified by the locus of \hi\ \citep{Shane:72} and 
 CO \citep{Cohen:80, Dame:86, Dame:01} emission in longitude-velocity ($l-v$) diagrams.
In molecular emission, \eg CO, the dominant large-scale feature in the inner Galaxy is known 
 as the Galactic Molecular Ring (GMR;  \citealt{Scoville:75, Clemens:88, Dame:01}) at Galactocentric radii
 between 4 and 7 kpc.
In a face-on deprojection, the GMR seems to consist of two spiral arms, the Sagittarius and Scutum
 arms \citep{Nakanishi:06}; the Scutum arm constitutes the inner side of the GMR.  
In the mid-infrared, a clear excess in star counts is seen toward the Scutum and Centaurus tangency
 directions, showing the Scutum arm is associated with overdensities with the old stellar disk
 \citep{Benjamin:09, Churchwell:09} and leading to the proposal that the Scutum and Perseus arms are
 the dominant, density-wave, arms of the Milky Way, extending symmetrically from opposite ends of the
 Galactic bar \citep{Drimmel:00, Benjamin:09, Churchwell:09}.
Recently, \cite{Dame:11} reported detection of a spiral arm in the first Galactic quadrant, $\sim 15$~kpc
 from the Galactic center, which might be the continuation of Scutum arm as a symmetric counterpart of
 the nearby Perseus arm.
  
While the locus of a spiral arm can be clearly traced in $l-v$ diagrams,
 a face-on deprojection of the arm is very uncertain, largely owing to very
 uncertain kinematic distances.  Kinematic distances in the direction of the Scutum arm
 at low galactic longitude are particularly uncertain due to the distance degeneracy
 for near-zero radial velocities (\eg \citealt{Sato:10G}).

In this paper, we report measurements of trigonometric parallaxes and proper motions of 
 \meth and \hho maser sources in six HMSFRs associated with the Scutum spiral arm.
Combined with ten other BeSSeL Survey parallaxes \citep{Xu:11, Immer:13, Zhang:13b},
 we now have a substantial sample of sources in the Scutum arm with accurate distances
 and motions, which enables a detailed study of the arm's location, pitch angle,
 and the peculiar motion of its sources.  

\section{OBSERVATIONS}\label{sect:obs}

We observed two 12-GHz \meth and four 22-GHz \hho maser sources in six Scutum HMSFRs.
Prior to our trigonometric parallax observations, we performed a VLBA survey of compact 
 extragalactic sources that may be used as position references for targeted maser source \citep{Immer:11}.
Also, accurate maser positions for VLBI observations were obtained in our preparatory survey 
 under VLBA program BR145A.

Multi-epoch phase-referenced observations were conducted under VLBA programs BR129B 
 (between 2007 October and 2009 March) and BR145I, M, O, Q, R and U
 (between 2010 September and 2012 September),
 to measure positions of the target sources relative to background quasars.
Table~\ref{tab:obs} lists the target masers and the observation dates.
For \meth masers, four observing epochs were selected to optimally sample the peaks
 of the sinusoidal parallax signature in right ascension over one year, maximizing the 
 sensitivity of parallax detection and ensuring that the parallax and proper motion signatures 
 are uncorrelated.  Since 22-GHz \hho masers are known to have short lifetimes ($\lax1$ yr), 
 we observed these masers six times in one year in a sequence that 
 was optimized for a parallax measurement provided a maser spot is detected for a span of 
 at least seven months.

Our observations were conducted as described by \cite{Reid:09I}.
In the phase-referenced observations, we employed four adjacent, dual circularly polarized, bands 
 of 8 MHz bandwidth for all sources, except for G025.70$+$00.04 where we used 4 MHz bandwidths.
The data correlation was performed at the VLBA correlation facility in Socorro, NM, using
 the DiFX software correlator\footnote[5]
{This work made use of the Swinburne University of Technology software correlator,
 developed as part of the Australian Major National Research Facilities Programme
 and operated under license.}
 \citep{Deller:07}.
The data were cross-correlated with an integration time of 0.9~seconds.
We correlated the data in two passes, resulting in frequency channels of 31.25 kHz, corresponding to velocity channels of 
 0.77 \kms\ and 0.42 \kms\ for the 12-GHz \meth and 22-GHz \hho maser emission, respectively,
 assuming rest frequencies of 12.178597 GHz for the \meth $2_{0} \rightarrow 3_{-1}$E
 and 22.235080 GHz for the \hho $6_{16}   \rightarrow 5_{13}$ transitions.
For the 12-GHz \meth maser source G025.70$+$00.04, the spectral and velocity resolutions 
 were 15.62 kHz and 0.38 \kms, respectively.
 
\section{DATA ANALYSIS AND RESULTS}\label{sect:results}

The correlated data were calibrated using the NRAO Astronomical Image
 Processing System (AIPS; \citealt{Greisen:03}) using scripts based on the ParselTongue package \citep{Kettenis:06},
 following the procedures described by \cite{Reid:09I}.
After calibration, we imaged the target maser and background continuum sources with the AIPS task IMAGR.
Then, we fitted elliptical Gaussian brightness distributions to the images to measure 
 the positions of the maser spots and background sources using the AIPS tasks JMFIT and SAD
 (more details of imaging and position fitting are described in \citealt{Sato:10W}).

\begin{deluxetable}{lrrclrrrl}
\tabletypesize{\scriptsize}
\tablecaption{Parallaxes and Proper Motions of High-Mass Star-Forming Regions in the Scutum Arm}
\tablewidth{0pt}
\tablehead{
\colhead{Source} & \colhead{$\ell$} & \colhead{\it{b}} & \colhead{Parallax} & \colhead{Distance} &
\colhead{$\mu_{x}$}  & \colhead{$\mu_{y}$} & \colhead{V$_{lsr}$} & \colhead{Ref.} \\
 & \colhead{(deg)} & \colhead{(deg)}  & \colhead{(mas)} & \colhead{(kpc)} & \colhead{(mas yr$^{-1}$)} & \colhead{(mas yr$^{-1}$)} &\colhead{(km s$^{-1}$)}
}
\startdata
\bf G005.88$-$00.39  &   5.89 & $-$0.40 & 0.334 $\pm$ 0.020 & $ ~2.99^{+0.19}_{\,-0.17} $ & 0.18 $\pm$ 0.34        & $-$2.26 $\pm$ 0.34 & 9 $\pm$ ~3 & 1, 5\\
\bf G011.91$-$00.61  & 11.92 & $-$0.61 & 0.297 $\pm$ 0.031 & $ ~3.37^{+0.39}_{\,-0.32} $ & 0.66 $\pm$ 0.28       &  $-$1.36 $\pm$ 0.41 &   37 $\pm$  ~5 &1 \\
G012.68$-$00.18       &  12.68 & $-$0.18 & 0.416 $\pm$ 0.028& $ ~2.40^{+0.17}_{\,-0.15} $ & $-$1.00 $\pm$ 0.95 &  $-$2.85 $\pm$ 0.95 &   58 $\pm$ 10 & 2  \\
G012.80$-$00.20       & 12.81 & $-$0.20  & 0.343 $\pm$ 0.037 &$ ~2.92^{+0.35}_{\,-0.28} $ & $-$0.60 $\pm$ 0.70 &  $-$0.99 $\pm$ 0.70 &   34 $\pm$  ~5 & 2\\
G012.88$+$00.48      & 12.89 & $+$0.49 & 0.400 $\pm$ 0.040 &$ ~2.50^{+0.28}_{\,-0.23} $ & 0.15 $\pm$ 0.25      &  $-$2.30 $\pm$ 0.39 &   31 $\pm$  ~7 & 2, 3 \\
G012.90$-$00.24       & 12.90 & $-$0.24  & 0.408 $\pm$ 0.025 &$ ~2.45^{+0.16}_{\,-0.14} $ & 0.19 $\pm$ 0.80      &  $-$2.52 $\pm$ 0.80 &   36 $\pm$ 10 & 2 \\
G012.90$-$00.26       & 12.91 & $-$0.26  & 0.396 $\pm$ 0.032 &$ ~2.53^{+0.22}_{\,-0.19} $ & $-$0.36 $\pm$ 0.80 &  $-$2.22 $\pm$ 0.80 &   39 $\pm$ 10 & 2   \\
\bf G013.87$+$00.28 & 13.87 & $+$0.28 & 0.254 $\pm$ 0.024 &$ ~3.94^{+0.41}_{\,-0.34} $ & $-$0.25 $\pm$ 2.00 &  $-$2.49 $\pm$ 2.00 &   48 $\pm$ 10 &1   \\
\bf G016.58$-$00.05  & 16.58 & $-$0.05  & 0.279 $\pm$ 0.023 &$ ~3.58^{+0.32}_{\,-0.27} $ & $-$1.13 $\pm$ 0.34  & $-$2.59 $\pm$ 0.35 & 60 $\pm$ ~5 & 1 \\
\bf G025.70$+$00.04 & 25.71 & $+$0.04 & 0.098 $\pm$ 0.029 &$ 10.20^{+4.29}_{\,-2.33} $ & $-$2.89 $\pm$ 0.07 &  $-$6.20 $\pm$ 0.36 &   93 $\pm$  ~5 &1 \\
G027.36$-$00.16       & 27.36  & $-0.17$ & 0.125 $\pm$ 0.042 &$ ~8.00^{+4.05}_{\,-2.01} $ & $-$1.81 $\pm$ 0.11 &  $-$4.11 $\pm$ 0.27 &   92 $\pm$  ~3 & 3 \\
G029.86$-$00.04       & 29.86  & $-$0.04 & 0.161 $\pm$ 0.020 &$ ~6.21^{+0.88}_{\,-0.69} $ & $-$2.32 $\pm$ 0.11 &  $-$5.29 $\pm$ 0.16 &  100 $\pm$  ~3 & 4 \\
G029.95$-$00.01       &  29.96 & $-$0.02 & 0.190 $\pm$ 0.019 &$ ~5.26^{+0.58}_{\,-0.48} $ & $-$2.30 $\pm$ 0.13 &  $-$5.34 $\pm$ 0.13 &   98 $\pm$  ~3  & 4 \\
G031.28$+$00.06      & 31.28 & $+$0.06 & 0.234 $\pm$ 0.039 &$ ~4.27^{+0.85}_{\,-0.61} $ & $-$2.09 $\pm$ 0.16 &  $-$4.37 $\pm$ 0.21 &  109 $\pm$  ~3 & 4 \\
G031.58$+$00.07      & 31.58 & $+$0.08 & 0.204 $\pm$ 0.030 &$ ~4.90^{+0.85}_{\,-0.63} $ & $-$1.88 $\pm$ 0.40 &  $-$4.84 $\pm$ 0.40 &   96 $\pm$  ~5 & 4 \\
\bf G032.04$+$00.05 & 32.04 & $+$0.06 & 0.193 $\pm$ 0.008 &$ ~5.18^{+0.22}_{\,-0.21} $ & $-$2.21 $\pm$ 0.40  & $-$4.80 $\pm$ 0.40 & 97 $\pm$ ~5 & 1 \\  
\hline
\enddata
\tablecomments{{ Column 1 lists the source names, which may differ slightly from those used in previous papers, 
 in order to follow the IAU rules for naming sources with Galactic coordinates.}
 Columns 2 and 3 give galactic longitude and
 latitude, respectively. Columns 6 and 7 are proper motions in the eastward
 ($\mu_x$=$\mu_\alpha$cos $\delta$) and northward directions ($\mu_y$ =
 $\mu_\delta$), respectively. Column 8 lists local standard of rest (LSR)
 velocity components, { estimated to be consistent with the thermal and maser lines \citep[and references therein]{Reid:13}.} } \label{tab:sctpar}
\tablerefs{
(1) this paper; (2) \cite{Immer:13}; (3) \cite{Xu:11}; (4) \cite{Zhang:13b}; (5) \cite{Hunter:08}
}
\end{deluxetable}

\setcounter{figure}{0}
\begin{figure*}
\begin{center}
\subfigure{
         \resizebox{10.5cm}{!}
           {\includegraphics[angle=270]{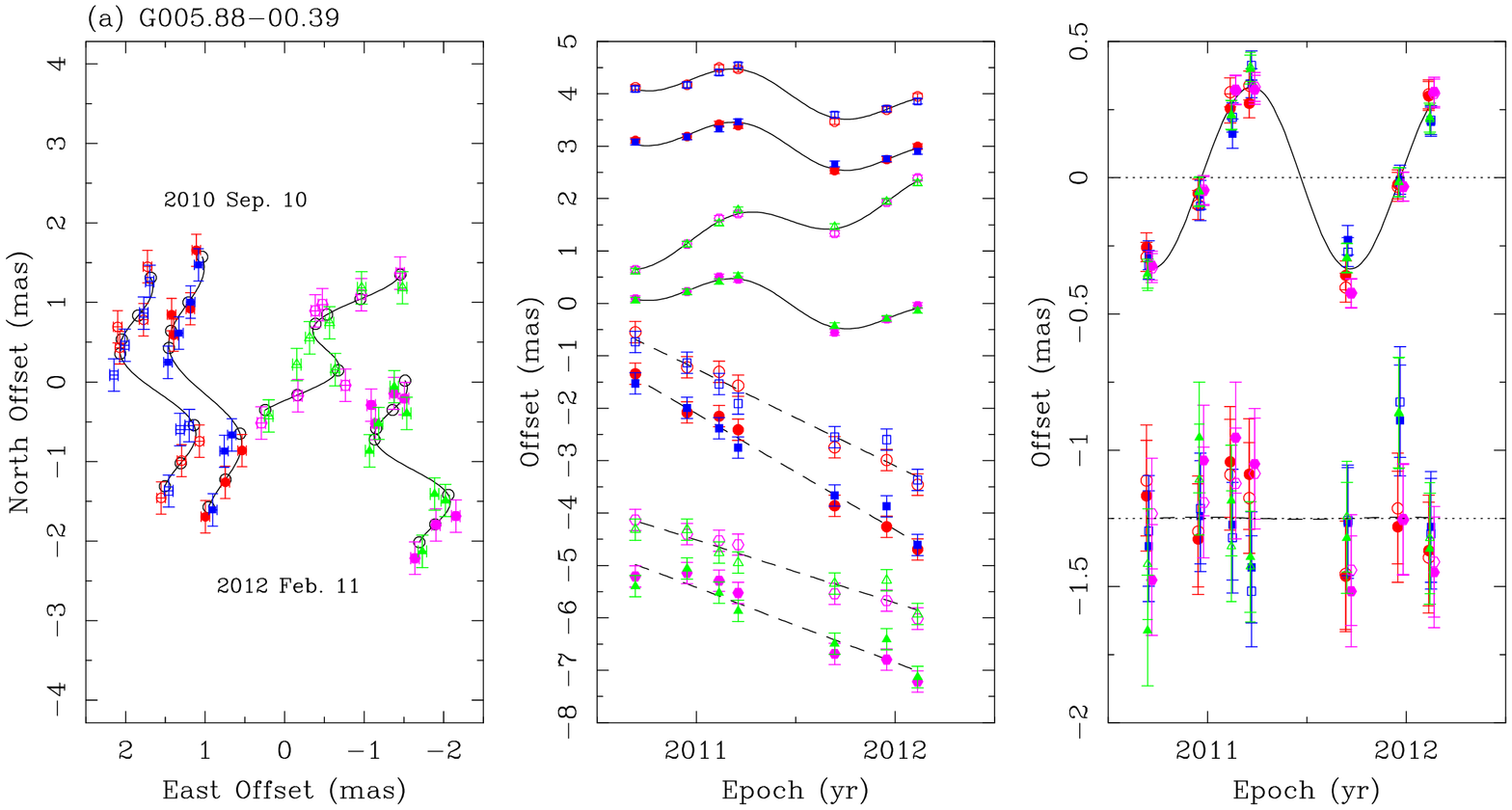}}
}
\subfigure{
         \resizebox{10.5cm}{!}
       {\includegraphics[angle=270]{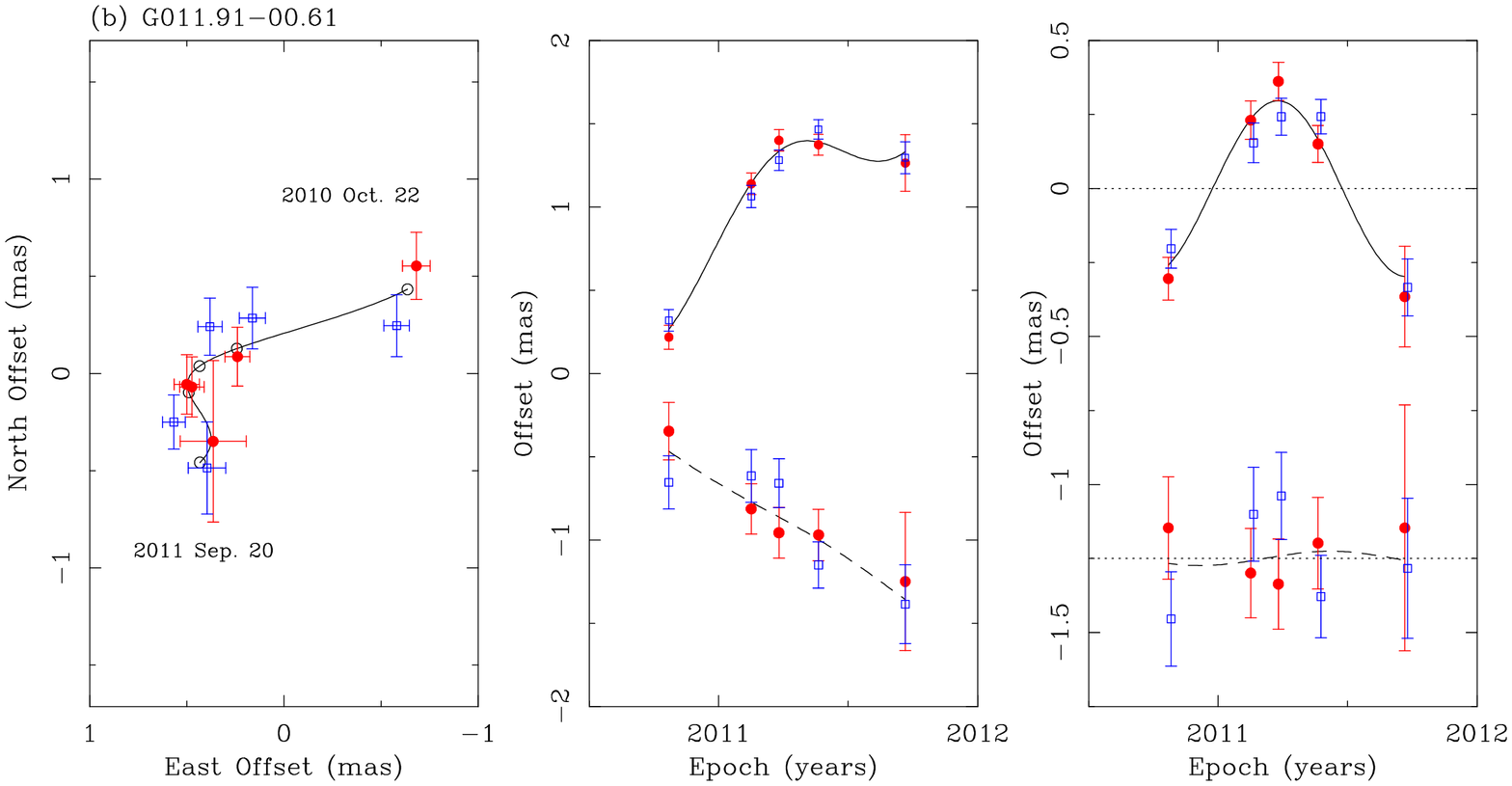}}
}
\subfigure{
         \resizebox{10.5cm}{!}{\includegraphics[angle=270]{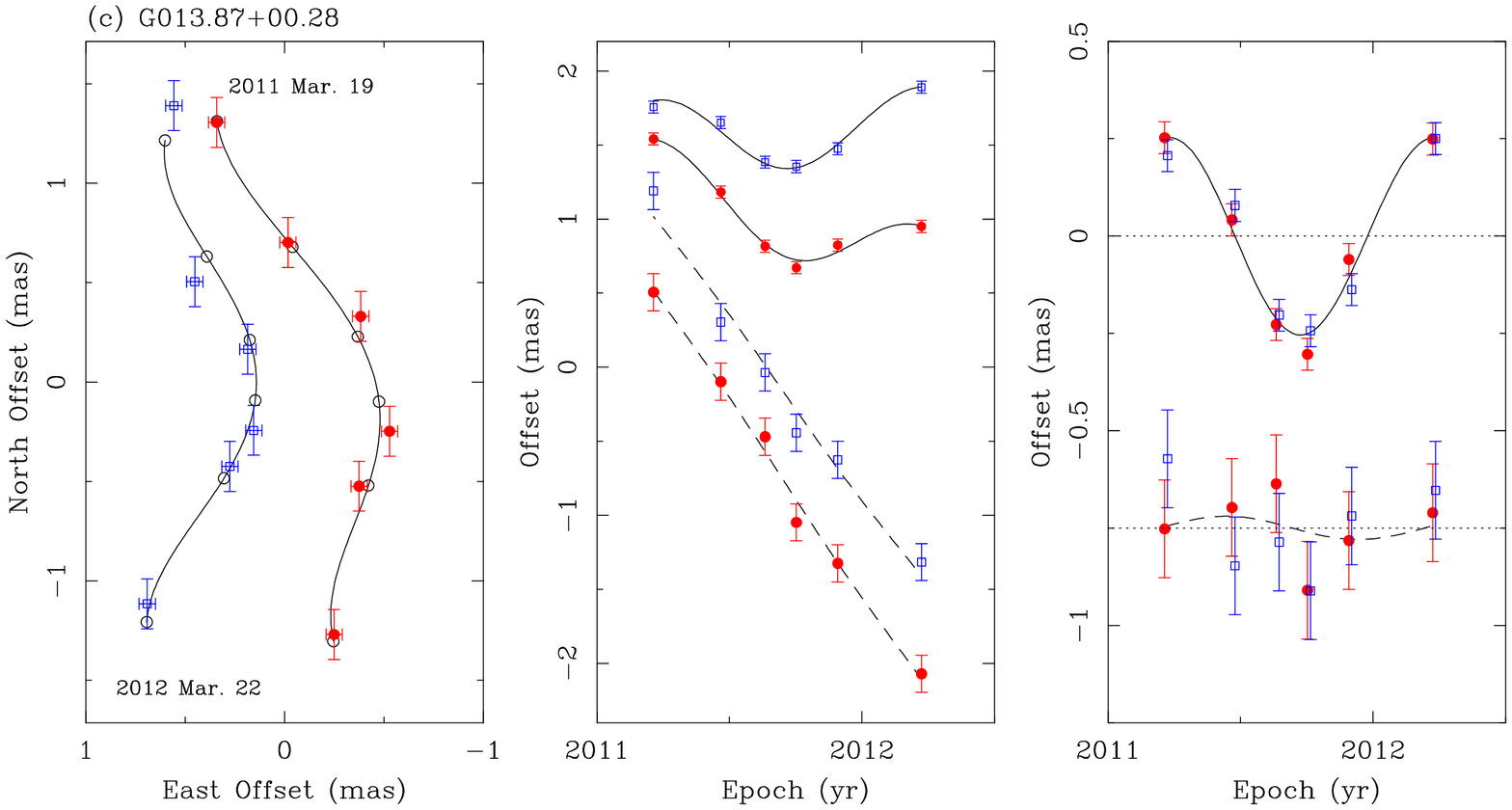}}
}
\caption{\scriptsize Parallax and proper motion data and fits for the observed maser sources.
Left panels: measured positions on the sky with first and last epochs labeled.
Data for different maser spots are offset for clarity.
The expected positions from the fit are indicated by open circles.
Middle Panels: east (solid lines) and north (dashed lines) position offsets versus time.
For clarity, data for different maser spots are offset vertically, the 
 northward data have been offset from the eastward data, and 
 small time shifts have been added to the data.
Right Panels: Same as middle panels,
 but now with the proper motion (top half) removed, showing only the 
 parallax effect; and with the entire (proper motion and parallax) fit 
 removed (bottom half), showing the residuals.
 (a) G005.88$-$00.39: four maser features, observed against two background quasars. 
 The reference quasars are indicated by the colors and shapes of the graph markers: pink pentagons and red circles for
 J1807$-$2506; green triangles and blue squares for J1755$-$2232.
 The four maser features are indicated as follows.
 The first feature consists of two spots at \vlsr\ velocities of 6.9 and 7.3 \kms\ and is represented (at the average positions of the two spots) 
 by open red circles and open blue squares.
 The second feature (filled red circles and filled blue squares) consists of spots at $\vlsr = 9.0$ and 9.4 \kms.
 The third feature (open pink pentagons and open green triangles) is made up of spots at $\vlsr = 10.3$ and 10.7 \kms.
 Finally, the fourth feature is at $\vlsr=11.1$ \kms\ and is represented by filled pink pentagons and filled green triangles. 
 (b) G011.91$-$00.61: one maser spot with $\vlsr=39.1$ \kms, measured relative to the two background 
 sources: J1808$-$1822 (red filled circles) and  J1825$-$1718 (blue open squares).
 (c) G013.87$+$0.28: two maser features consisting of three spots each,
  both measured relative to a single background quasar (J1809$-$1520). 
 One feature with \vlsr\ between 15.0 and 16.7 \kms\ (red filled circles) and 
 the other with \vlsr\ between 17.5 and 18.4 \kms\ (blue open squares).}
\label{fig:parallaxes}
\end{center}
\end{figure*}

\addtocounter{figure}{-1}
\addtocounter{subfigure}{3}
\begin{figure*}
\begin{center}
\subfigure{
         \resizebox{10.5cm}{!}{\includegraphics[angle=270]{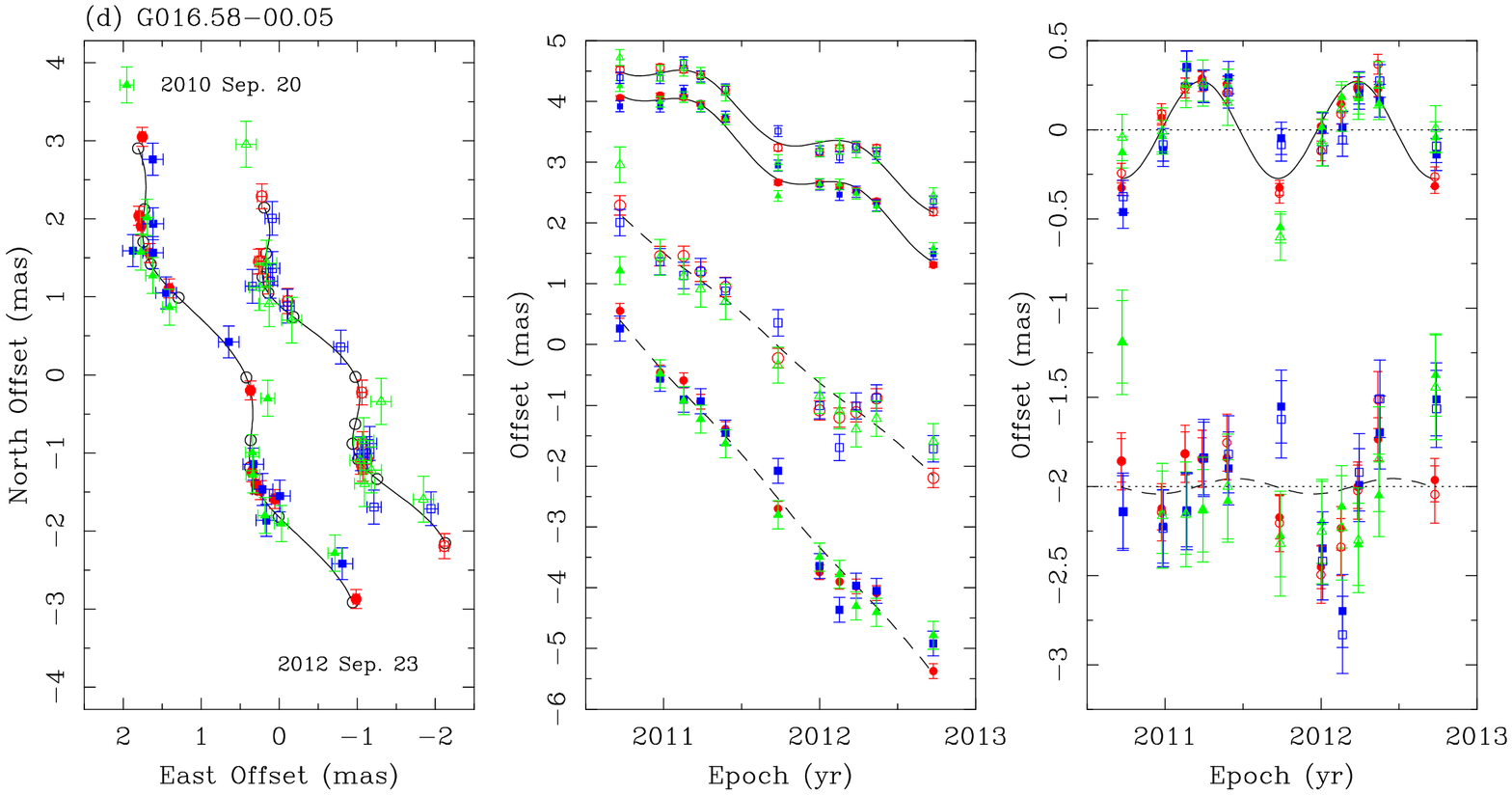}}
}
\subfigure{
         \resizebox{10.5cm}{!}{\includegraphics[angle=270]{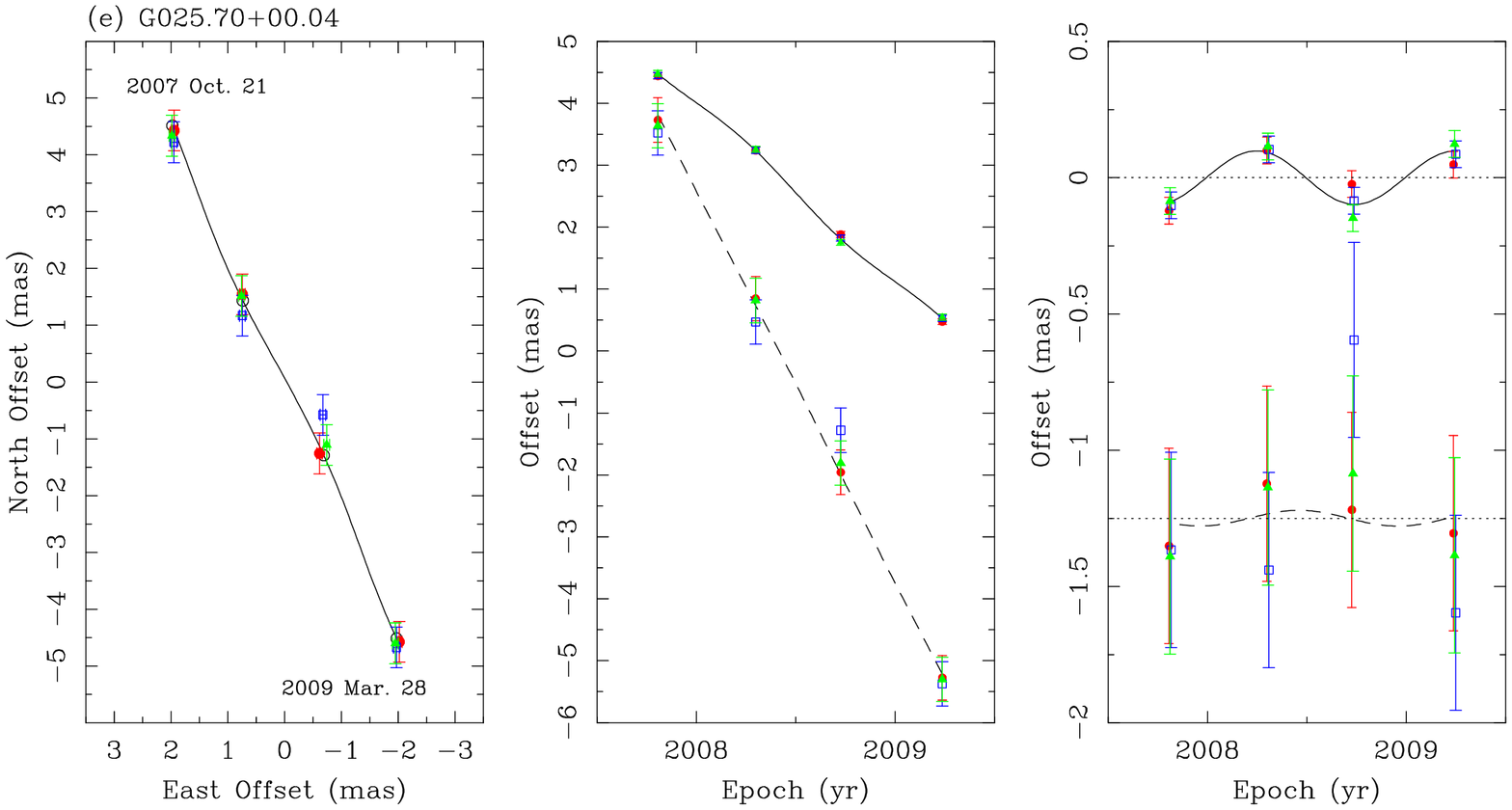}}
}
\subfigure{
         \resizebox{10.5cm}{!}{\includegraphics[angle=270]{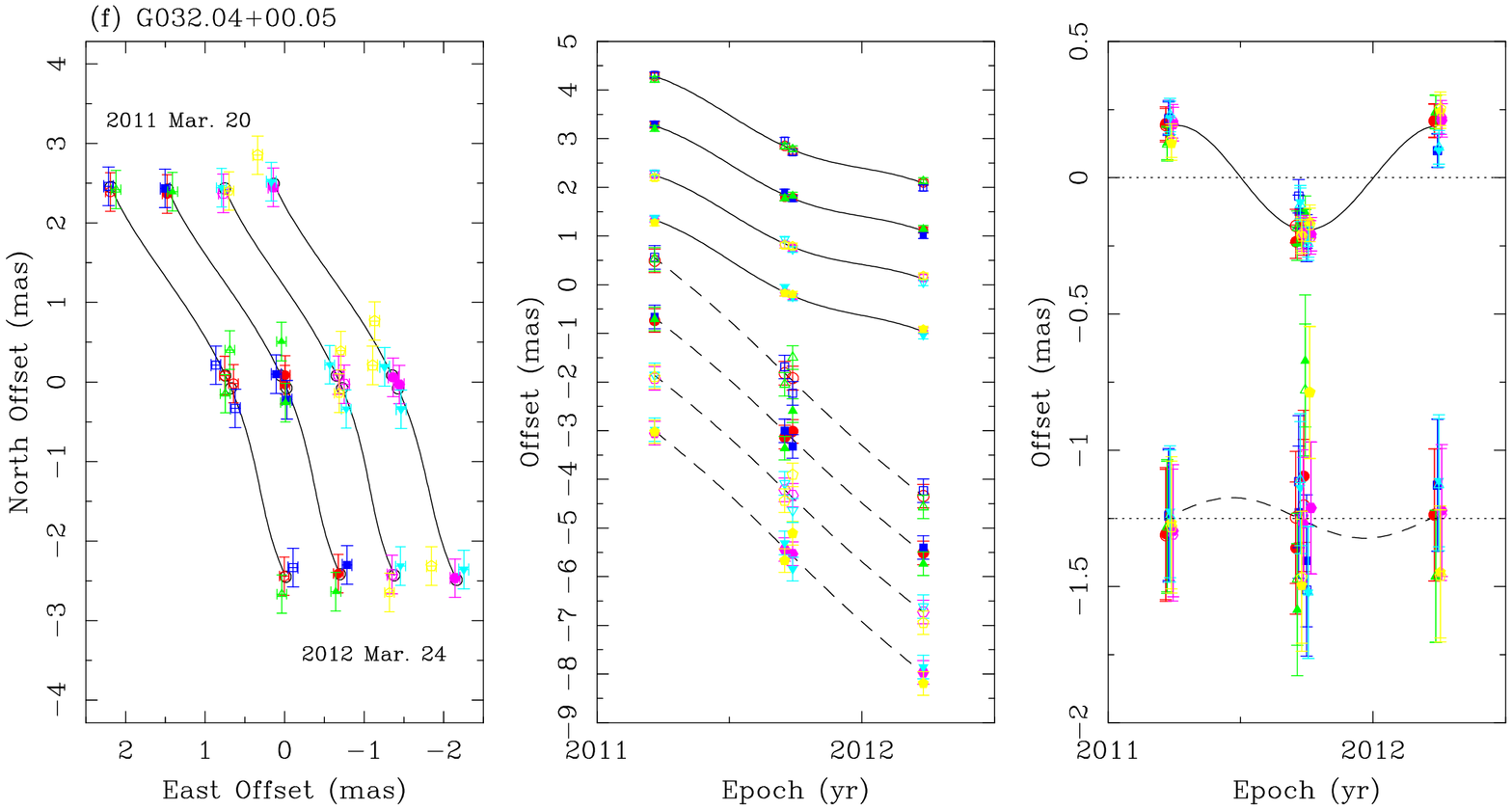}}
}
\caption{\footnotesize
        {\it Continued.}
(d) G016.58$-$00.05: two features: one with spots at \vlsr\ 
between 63.4 and 64.3 \kms\ (filled symbols) and 
the other with \vlsr\ between 62.6 and 63.0 \kms\ (open symbols). 
For this source, three reference quasars were used, namely
 J1825$-$1551 (red circles), J1825$-$1718 (blue squares), and J1809$-$1520 (green triangles).
(e) G025.70$+$00.04: one maser feature consistent of three spots, measured
against a single background quasar (J1837$-$0653). The spots are as follows:
one spot with $\vlsr=89.6$ \kms\ (red filled circles), a second spot with 
$\vlsr=94.6$ \kms\ (blue open squares) and a 
third spot with $\vlsr=95.4$ \kms\ (green filled triangles).
(f) G032.04$+$00.05: four maser features, measured against three background sources.
The background quasars are J1857$-$0048 (pink hexagons and red circles), 
J1904$+$0110 (cyan inverted triangles and blue squares) and 
J1833$-$0323 (yellow pentagons and green triangles). 
The first feature is at $\vlsr=101.1$ \kms\ (filled pink, cyan and yellow markers);
the second feature has $\vlsr=100.3$ \kms\ (open pink, cyan and yellow markers);
the third features has $\vlsr=99.5$ \kms\ (filled red, blue and green markers); 
and the final feature consists of two spots, at $\vlsr=98.8$ and 98.0 \kms\ 
(open red, blue and green markers).
}
\end{center}
\end{figure*}

The maser positions relative to the background sources were modeled with
 a sinusoidal parallax signature and linear proper motions in each coordinate. 
In order to estimate the systematic errors, we added ``error floors'' in quadrature to the 
 formal position uncertainties (separately for the right ascension and declination data)
 and adjusted them to yield post-fit residuals with $\chi^2$ per degree of freedom near unity 
 for both coordinates (see \citealt{Reid:09I}), except for G016.58$-$00.05 (see Section~\ref{sect:g16}).

As a conservative approach to estimating the parallax uncertainties, 
 we multiplied the formal least-square error by a factor of $\sqrt{N_{\rm spot}}$,
 where $N_{\rm spot}$ is the number of maser spots used to fit the parallax.
This assumes that each spot position is 100\% correlated with the other spot positions,
 due to systematic position errors independent of intrinsic maser structure.
The dominant source of position error is uncompensated atmospheric delay differences between 
 the target maser and a background source, which equally affects all maser spot positions for the same epoch.
Using background sources at different positions on the sky reduces this error, and therefore,
 we assume no correlation for data from different background sources.

Table~\ref{tab:sctpar} lists the results of the parallax fitting for the six HMSFRs of this study;
Figure~\ref{fig:parallaxes} shows the parallax fit for each source.  Detailed discussion on each 
 source is given in Appendix~\ref{app:description}.

\section{DISCUSSION}\label{sect:tint}

\subsection{Geometry of the Scutum arm} 

\subsubsection{Arm Assignment} 

Since the first 21 cm surveys of the Galaxy (\eg \citealt{Westerhout:57}),
 quasi-continuous emission features have been repeatedly identified
 in $l-v$ diagrams as spiral arms: the Outer Arm, Perseus, Sagittarius-Carina,
 Scutum-Centaurus, Norma-4kpc, and the Expanding 3-kpc arm.  
The same features emerged with greater contrast in Galactic CO surveys (\eg \citealt{Cohen:80}).  
However, attempts to map Galactic spiral structure from these $l-v$ features have been stymied by
 uncertainties in transforming from velocity to (kinematic) distance.  
Parallax measurements provide ``gold standard'' distances, and with a sufficient number of 
 parallax measurements of HMSFRs, it may soon be possible to map in detail Galactic spiral 
 structure. 

\begin{figure*}
	\centering
   \includegraphics[width=14cm, angle=0]{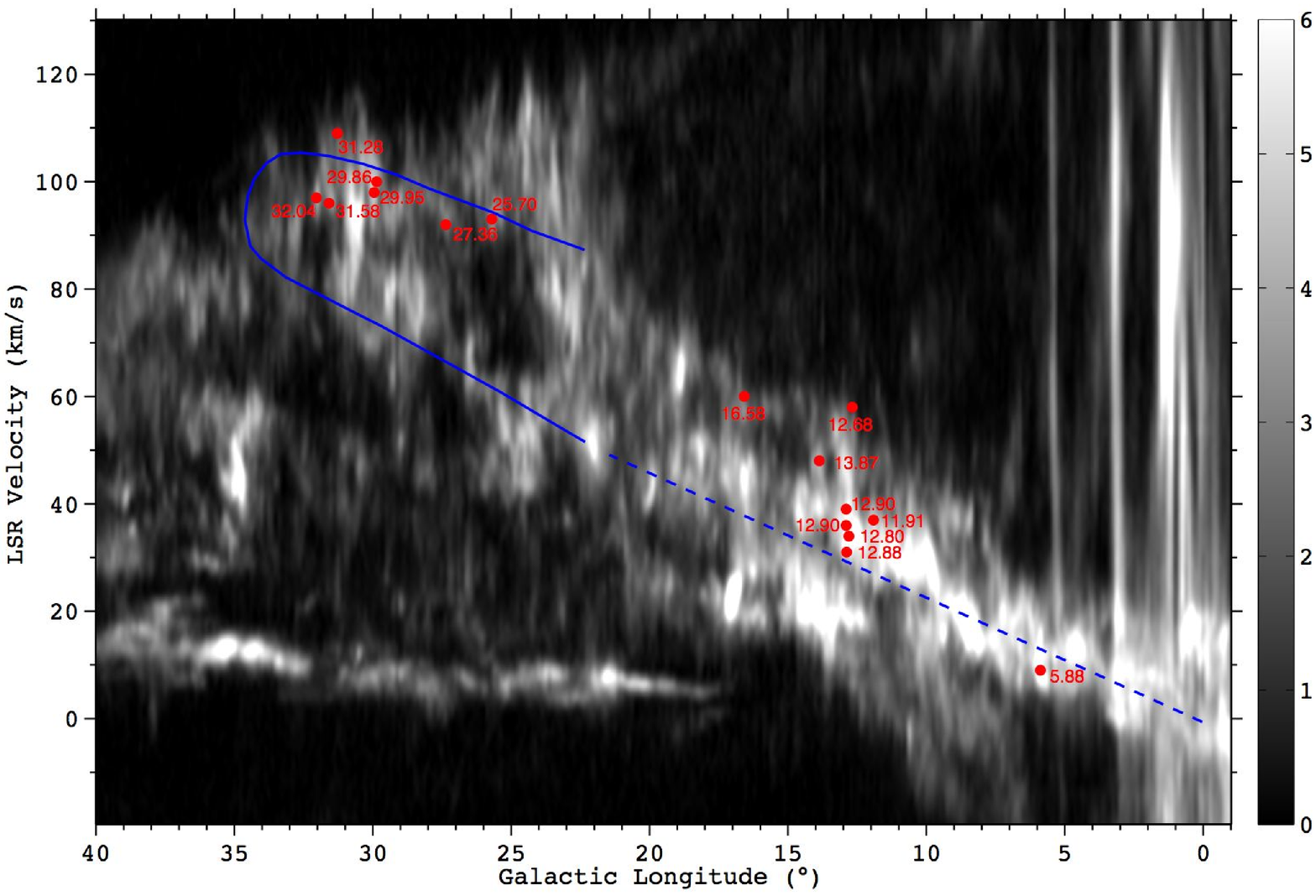}
		\caption{\footnotesize The locus of the Scutum arm in longitude and velocity (blue line)
		 overlaid on CO emission from \cite{Dame:01}.  The solid line is taken from
		 an early 21 cm study of the arm by \cite[his Fig. 2]{Shane:72}; 
		 the dotted line simply extrapolates that locus to the origin, through which it must pass 
		 in the absence of non-circular motions.  
		 The CO emission in integrated over $\pm 1^\circ$ of Galactic latitude,
		  and the grayscale runs from 0 K arcdeg (black) to 6 K arcdeg (white). 
		  Red dots mark the HMSFRs which we associate with the Scutum arm. 
		}\label{fig:lvarm}
\end{figure*}

Even a moderate number of parallax distances to HMSFRs, confidently assigned to an $l-v$ spiral 
arm, can provide extremely valuable and heretofore unattainable information on spiral arm location 
and shape.  We have therefore assigned our HMSFRs to spiral arms by first associating them with molecular 
clouds in the CO survey of \cite{Dame:01} or \cite{Jackson:06} and, in turn, associating the clouds
kinematically with a classic $l-v$ spiral feature.  
The $l-v$ ``loop'' of the Scutum arm, as identified in 21 cm emission by \cite{Shane:72} and in CO by \cite{Cohen:80},
is shown on a CO $l-v$ diagram in Fig.~\ref{fig:lvarm}, along with the HMSFRs associated with that arm. 
Generally, the spread of sources about the arm locus is consistent with a Virial spread
for the exciting stars with their molecular clouds, cloud-cloud velocity dispersion,
and, of course, uncertainty of the $l-v$ locus itself.  
The exception is G012.68$-$00.18, which \cite{Immer:13} identified as a probable velocity outlier
to the W33 region at a velocity of $\sim 35$~\kms. 
For G016.58$-$00.05, two CO peaks are found at velocities $\sim 59$~\kms\ and $\sim 46$~\kms,
 the latter corresponding to the Scutum arm. 
We assigned this source to the Scutum arm based on
 both the CO velocity and parallax distance.
In addition to the six HMSFRs of this study, we assigned 10 other BeSSeL/VLBA maser sources, 
whose trigonometric parallaxes have been reported by \cite{Xu:11}, \cite{Immer:13}, and \cite{Zhang:13b}, 
to the Scutum arm  (Table~\ref{tab:sctpar}).
We thus have a total of 16 HMSFRs with trigonometric parallaxes for the Scutum arm, which
enables us to examine the geometry of the inner part of the spiral arm with a substantial 
sample. 

\begin{figure}
   \includegraphics[width=7cm, angle=0]{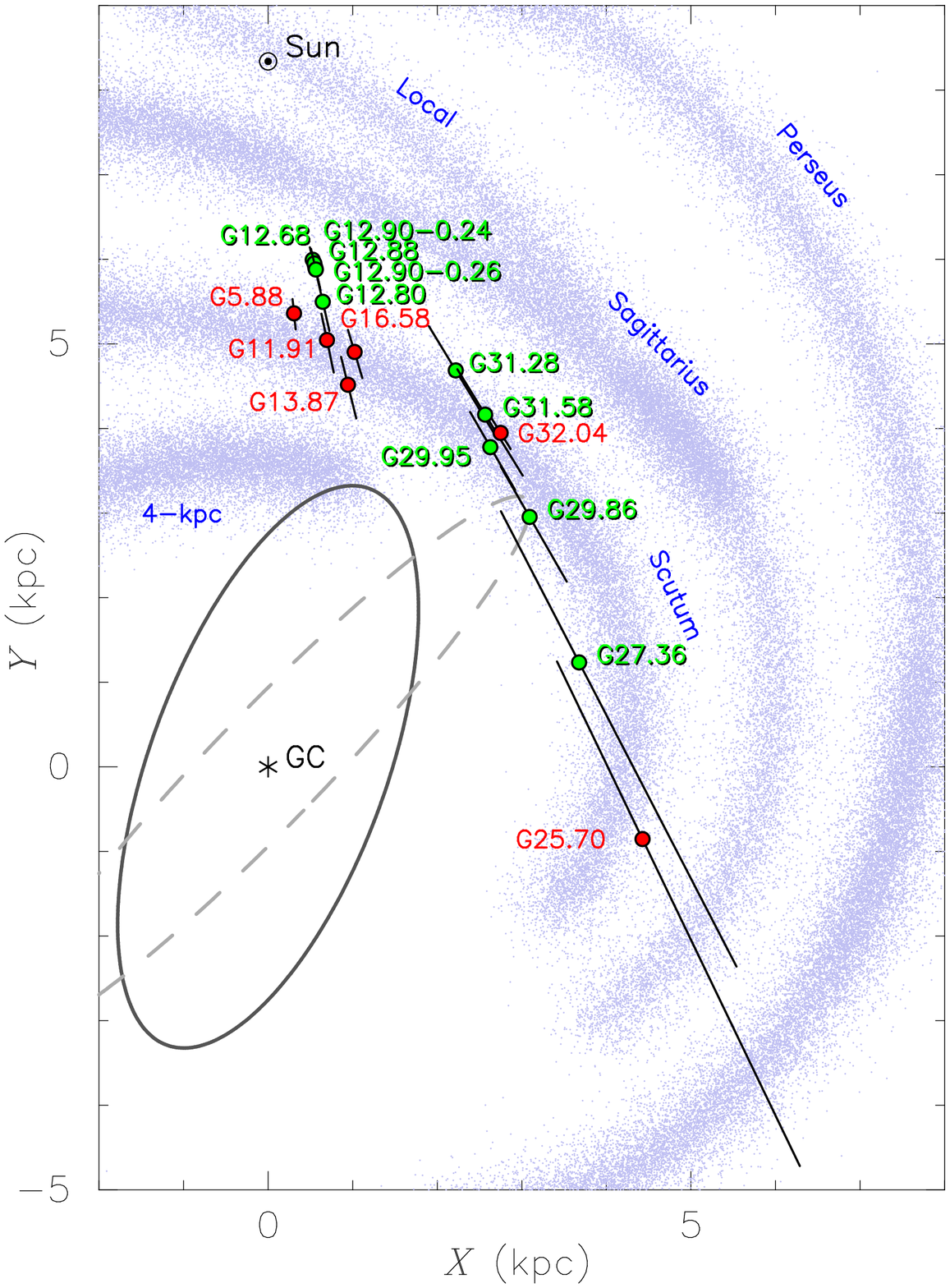}
		\caption{\footnotesize Locations of 16 HMSFRs in the Scutum arm in the Milky Way for which trigonometric parallaxes have been measured.
		1$\sigma$ distance error bars are indicated.
	          Six HMSFRs from this study are shown by red dots and labeled with the galactic longitudes.
	          Ten other HMSFRs from BeSSeL/VLBA papers are marked by green dots (Table~\ref{tab:sctpar}).
	          { $R_0 = 8.34$~kpc } was adopted \citep{Reid:13}.
	          The positions of the Galactic center (GC) and the Sun are marked by an asterisk ($*$) and $\odot$, respectively.
	          The background is the spiral arm model of the Milky Way by \cite{Taylor:93} 
	           with a slightly modified (more axisymmetric) curvature in the near part of the Sagittarius arm. 
	           Spiral arm names are labeled in blue.
	           The grey ellipses show the locations of the central bar (solid curve) from \cite{Gerhard:02} and \cite{Churchwell:09}
	            and the long bar (dashed curve) from  \cite{Benjamin:05} in the Galactic center region.         
	          }\label{fig:gcplot}
\end{figure}

Table~\ref{tab:sctpar} lists the measured parallaxes of each star-forming region and the 
estimated proper motion of the central star.  Figure~\ref{fig:gcplot} shows the locations of all 
these sources in the Milky Way. 
Note that the 4 most distant sources in Fig.~\ref{fig:gcplot}
 (G029.95$-$00.01, G029.86$-$00.04, G027.36$-$00.16, and G025.70$+$00.04)
 are the same ones that would be expected to be most distant based on their $l-v$ locations
 in the upper (far-side) section of the Scutum loop in Fig.~\ref{fig:lvarm}.
Even their relative distances are correct, since the distance to the far side should increase with decreasing longitude. 
This agreement supports the $l-v$ locus of Scutum proposed by \cite{Shane:72} and others subsequently,
 and suggests that the arm continues for some distance past its tangency near the end of the central bar,
 { as recently shown by \citet{Ellsworth-Bowers:13}.}

\subsubsection{Pitch Angle} 

\begin{figure*}
	\centering
   \includegraphics[width=13cm, angle=0]{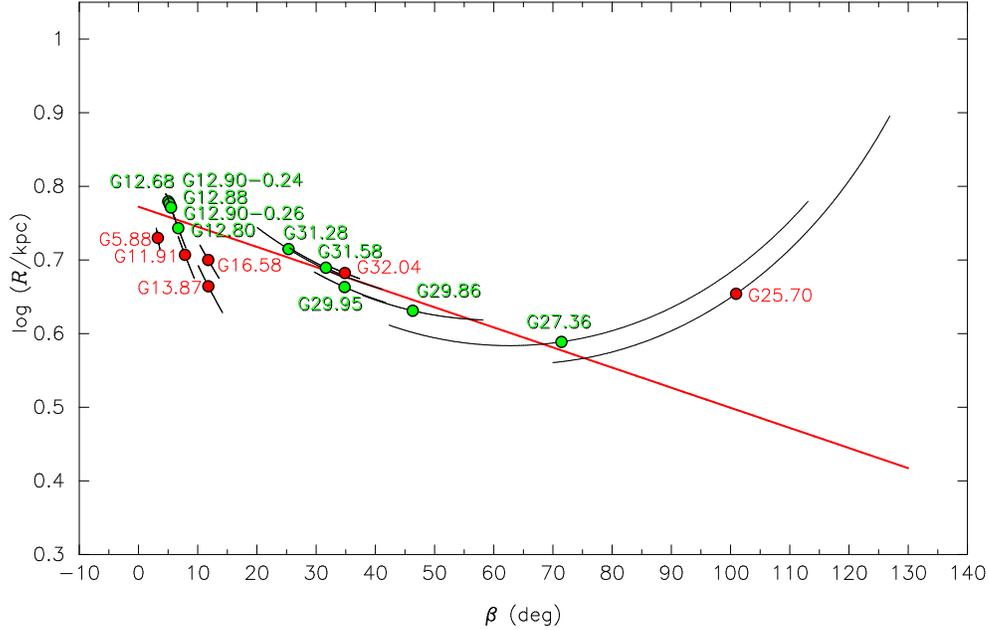}
		\caption{\footnotesize Pitch angle of the Scutum arm.  The logarithm of Galactocentric radius ($R$) in kpc units
		is plotted against Galactocentric azimuth ($\beta$), for $R_0$ of { 8.34 kpc}. 
		A total of 16 HMSFRs in the Scutum arm with parallax measurements are plotted with 
		 1$\sigma$ parallax uncertainties (black curves): red dots for 6 sources from this study and	 
	         green dots for 10 sources from other BeSSeL/VLBA papers (Table~\ref{tab:sctpar}).
		Red solid line shows the straight line (spiral arm) fitted by a Bayesian procedure to minimize distance of each source perpendicular to the line (see text).
		}\label{fig:pitch}
\end{figure*}

For an ideal log-periodic spiral arm, the arm geometry can be defined by
$\ln (R/R_{\rm ref}) = - (\beta - \beta_{\rm ref}) \tan \psi$ 
in a Galactocentric reference frame, where $R$ is the radius from the Galactic center,
$\beta$ is the Galactocentric azimuth (0 toward the Sun and increases with galactic longitude), 
and $R_{\rm ref}$ is the radius at a reference longitude of $\beta_{\rm ref}$ \citep{Reid:09VI}.
The pitch angle $\psi$ of the arm is constant for a logarithmic arm, defined as the angle
between the spiral arm and a tangent to a Galactocentric circle at the same radius.
In a plot of the logarithm of Galactocentric radius, $R$ (in kpc units), against Galactic azimuth, 
$\beta$, a log-periodic spiral arm section is a straight line, with the slope of the line 
proportional to the tangent of the pitch angle, $\psi$. 
In Fig.~\ref{fig:pitch}, we plot log~($R/1$kpc) versus $\beta$ for the Scutum arm, in the same 
manner as \cite{Reid:09VI} did for three other spiral arms.  
We used a Bayesian fitting procedure 
to minimize the distance of each source perpendicular to the fitted straight line.
We adopted the distance to the Galactic center of { $R_0 =  8.34$~kpc} \citep{Reid:13}.
{ In determining the pitch angle, the only assumption is the log-periodic shape of the spiral arm. 
No kinematic assumptions (such as circular rotation) were made.
The assumption of a log-periodic shape is well supported by observations of other galaxies. 
Admittedly pitch angles of spiral arms in other galaxies often vary in different portions of an arm,
 implying that a non-logarithmic model could also be possible, 
 but the present data does not justify a model of such complexity.}

{ The deviations of source locations from a simple logarithmic spiral model,
 beyond those expected from parallax uncertainty, are expected from a non-zero arm width (thickness).
As an approximation of the finite width of the spiral arm,
 we added an ``astrophysical noise'' in quadrature with measurement uncertainty
 and adjusted its magnitude to obtain a reduced $\chi_\nu^2 \approx 1$. }

We find a pitch angle of { $19\d8\pm 3\d1$} for this arm and 
 $\ln(R_{\rm ref}/{\rm kpc}) = 1.62\pm0.02$ at $\beta_{\rm ref}=25.5^\circ$
 (uncertainties give 68\% confidence ranges).  
We added an astrophysical noise of { 0.17 kpc} (to obtain $\chi_\nu^2 \approx 1$). 
This pitch angle is larger than the pitch angles of $6^\circ-12^\circ$ for other spiral arms of 
the Milky Way, recently estimated with BeSSeL Survey results \citep{Sanna:12, Zhang:13a, Xu:13, 
Wu:13}. 
This pitch angle is also larger than the $14\d2$ derived by \cite{Dame:11} for the Scutum arm using 
a wider range of Galactocentric azimuths, based on a CO $l-v$ diagram of the first 
and fourth Galactic quadrants.
{ As a test, we removed the most distant (and uncertain) source (G025.70$+$00.04) 
 and re-fitted for pitch angle; this increased the estimated pitch angle ($21\d7\pm 3\d5$). 
}
The large pitch angle we obtain for the inner part of the Scutum arm
 may be due to the source locations near the bar in the Galactic center region. 
{ Arms coming off of a bar in other galaxies can display large (and sometimes varying) curvature.}
Measurements of spiral arm pitch angles in external galaxies by \cite{Davis:12} 
 show that the barred geometry of a spiral galaxy can significantly bias the pitch angle values 
 at inner radii toward higher values than the global pitch angle of the spiral arms.
Among the 16 HMSFRs in the Scutum arm, G029.86$-$00.04 is likely located close to the
 near end of the long bar \citep{Hammersley:00, Benjamin:05}
 and G027.36$-$00.16 and G025.70$+$00.04 may lie at Galactocentric radii
 smaller than the semi-major axis of the long bar (see Fig.~\ref{fig:gcplot}).
More parallax measurements are at a wider range of Galactocentric azimuths are required
 to reliably measure the global pitch angle of the spiral arm.

\subsection{3-D Motions} 

Using the measured parallaxes and proper motions of the 16 HMSFRs in the Scutum arm,
along with radial velocities and source coordinates, we can calculate 3-dimensional motions of the 
sources in this arm with respect to the rotation of the Galactic disk.
Table~\ref{tab:sctpar} lists measured values for parallaxes and systemic motion components
for all sources.  By transforming to a reference frame that rotates with the Galaxy, 
we obtain the source peculiar motion components ($U_s$, $V_s$, $W_s$),
where $U_s$ is toward the Galactic center, $V_s$ is in the direction of Galactic rotation,
$W_s$ is toward the north Galactic pole (NGP) at the locations of the sources.
 
\begin{figure}
   \includegraphics[width=7cm, angle=0]{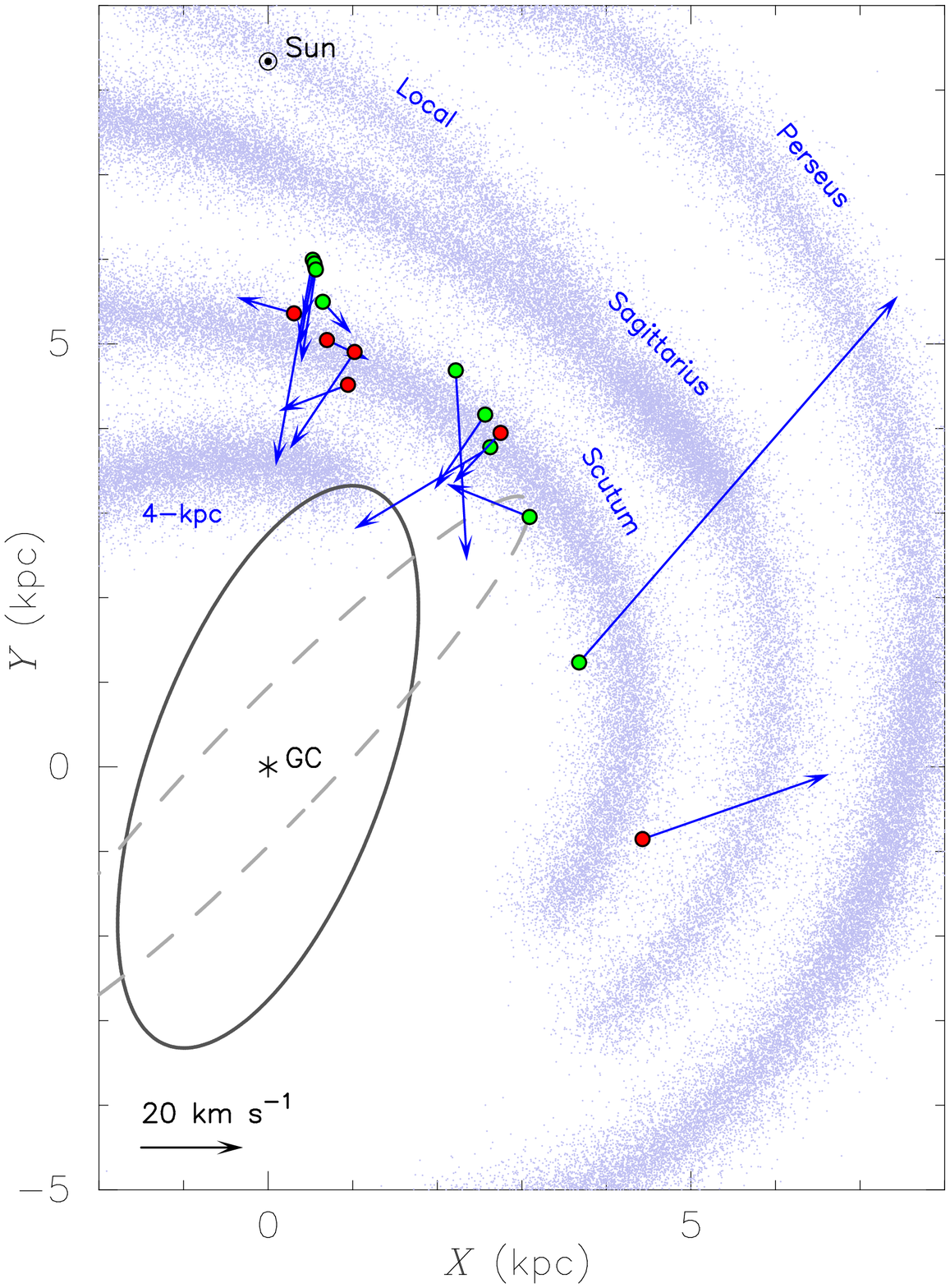}
		\caption{\footnotesize Peculiar motions of 16 HMSFRs in the Scutum arm in the Milky Way relative to 
		the Galactic rotation model { A5} from \cite{Reid:13}, described in the text.
		As in Fig.~\ref{fig:gcplot}, red and green dots show six HMSFRs from this study and 10 other HMSFRs from BeSSeL/VLBA papers, respectively.
		Blue arrows indicate the motion of each region relative to the Galactic rotation model, as listed in Table~\ref{tab:sctmot}.
		A sample magnitude for a motion of 20 \kms\ is shown at the left bottom.
	          }\label{fig:sctmot}
\end{figure}

Table~\ref{tab:sctmot} shows the peculiar motions of the sources relative to the Galactic rotation 
 model { ``A5''} of \citet{Reid:13}, { which is based on 80 parallaxes of star forming regions across 
 the Galaxy from the BeSSeL and VERA results:}
 { $R_0=8.34\pm 0.16$~kpc, $\Theta_0 = 240 \pm 8$~\kms\ and $\rm{d}\Theta/\rm{d}R = -0.2\pm 0.4$~\kms~kpc$^{-1}$.} 
The adopted Solar motion is
 { $U_\odot = 10.7 \pm 1.8$~\kms, $V_\odot = 15.6 \pm 6.8$~\kms, and $W_\odot = 8.9 \pm 0.9$~\kms.}
The peculiar motions of the Scutum arm sources relative to this Galactic rotation model are 
 plotted in Fig.~\ref{fig:sctmot}.  
The uncertainties of the peculiar motions were calculated numerically.  
We included 7~\kms\ virial component ``noise'' for each source (and excluded trial distances beyond 
12 kpc).
For the Scutum arm sources, we find 
{ $\bar{U_s} = 8.2 \pm 2.3$~\kms, $\bar{V_s} = -4.2 \pm 2.6$~\kms, and $\bar{W_s}=-1.5 \pm 1.5$~\kms.} 

\begin{deluxetable}{lrrrrr}
\tabletypesize{\scriptsize}
\tablecaption{Motions of High-Mass Star-Forming Regions in the Scutum Arm}
\tablewidth{0pt}
\tablehead{
 \colhead{Source}  &\colhead{$R_s$} & \colhead{$\beta$} &  \colhead{$U_s$} & \colhead{$V_s$} & \colhead{$W_s$} \\
  &\colhead{(kpc)} & \colhead{(deg)} & \colhead{(\kms)} & \colhead{(\kms)}  &  \colhead{(\kms)} 
}
\startdata
\bf G005.88$-$00.39  &  $  5.37^{+ 0.23}_{\, -0.25}$ & $   3.3^{+ 0.4}_{\, -0.3}$     & $-$2.5 $\pm$ ~3.8      &   $-$11.5 $\pm$ ~8.5  &    $-$9.2 $\pm$ ~5.1 \\
\bf G011.91$-$00.61  &  $  5.09^{+ 0.33}_{\, -0.42}$ & $   7.8^{+ 1.9}_{\, -1.1}$     &       2.7 $\pm$ ~8.5      &           8.8 $\pm$ ~9.3  & $-$11.0 $\pm$ ~5.7  \\ 
G012.68$-$00.18       &  $  6.02^{+ 0.21}_{\, -0.23}$ & $   5.0^{+ 0.6}_{\, -0.4}$     &      41.1 $\pm$ 10.3     &  $-$3.7 $\pm$ 13.0     &         3.2 $\pm$ 10.9  \\
G012.80$-$00.20       &  $  5.54^{+ 0.30}_{\, -0.39}$ & $   6.7^{+ 1.5}_{\, -0.9}$     &       5.6 $\pm$  ~7.9     &        6.4 $\pm$ 11.9     &         9.6 $\pm$ ~9.9  \\
G012.88$+$00.48      &  $  5.93^{+ 0.26}_{\, -0.32}$ & $   5.4^{+ 1.0}_{\, -0.6}$     &     10.6 $\pm$ ~7.6      &  $-$1.4 $\pm$ ~8.7     &   $-$5.6 $\pm$ ~3.9  \\
G012.90$-$00.24       &  $  5.98^{+ 0.21}_{\, -0.22}$ & $   5.3^{+ 0.5}_{\, -0.4}$     &     16.3 $\pm$ 10.4      &   $-$1.4 $\pm$ ~11.6    &   $-$7.1 $\pm$ ~9.5  \\
G012.90$-$00.26       &  $  5.91^{+ 0.24}_{\, -0.27}$ & $   5.5^{+ 0.8}_{\, -0.5}$     &     18.5 $\pm$ 10.4      &    $-$1.1 $\pm$ 12.1    &   $-$0.1 $\pm$ ~9.8 \\
\bf G013.87$+$00.28 &  $  4.62^{+ 0.34}_{\,-0.42}$ & $  11.8^{+ 2.9}_{\, -1.6}$    &      ~7.9 $\pm$  20.8    & $-$12.3 $\pm$ 35.1    &   $-$9.0 $\pm$ 37.6  \\
\bf G016.58$-$00.05  &  $  5.01^{+ 0.28}_{\,-0.33}$ & $  11.8^{+ 2.1}_{\, -1.4}$    &     21.3 $\pm$  ~8.0     & $-$8.6 $\pm$ ~9.5    &         5.1 $\pm$ ~6.0  \\
\bf G025.70$+$00.04 & $  4.51^{+ 3.00}_{\,- 0.68}$ & $ 100.9^{+20.1}_{\,-28.5}$ & $-$33.7 $\pm$ 38.5  & $-$19.5 $\pm$ 21.5   & $-$4.5 $\pm$ ~8.2   \\
G027.36$-$00.16       &  $   3.88^{+ 1.99}_{\,- 0.08}$ & $  71.4^{+33.3}_{\,-25.4}$ & $-$82.5 $\pm$ 50.2  & $-$48.4 $\pm$ 27.4    & $-$1.6 $\pm$ ~6.4  \\ 
G029.86$-$00.04       &  $   4.28^{+ 0.25}_{\, -0.13}$ & $  46.3^{+13.3}_{\, -8.0}$ &       7.3 $\pm$ 22.8      & $-$16.0 $\pm$ ~7.8    & $-$1.5 $\pm$ ~4.0  \\
G029.95$-$00.01       &  $   4.60^{+ 0.26}_{\, -0.23}$ & $  34.8^{+ 7.9}_{\, -4.9}$  &       28.5 $\pm$ 14.5   & $-$12.8 $\pm$ ~7.5    & $-$0.9 $\pm$ ~3.6   \\
G031.28$+$00.06      &  $   5.19^{+ 0.38}_{\, -0.41}$ & $  25.3^{+11.2}_{\, -4.9}$ &       33.0 $\pm$ 22.1   &         18.0 $\pm$ ~9.7  &       6.3 $\pm$ ~3.8  \\
G031.58$+$00.07      &  $   4.89^{+ 0.34}_{\, -0.30}$ & $  31.7^{+11.5}_{\, -5.8}$ &       17.8 $\pm$ 21.8   & $-$0.8 $\pm$ ~10.6      & $-$3.4 $\pm$ ~9.9  \\
\bf G032.04$+$00.05 &  $   4.81^{+ 0.16}_{\, -0.15}$ & $  34.8^{+ 2.9}_{\, -2.4}$  &       13.5 $\pm$ 12.9   & $-$2.0 $\pm$ ~8.6      &       3.5 $\pm$ 9.9 \\
\hline
Error-weighted mean &    & &     8.2 $\pm$ ~2.3   & $-$4.2 $\pm$ ~2.6   & $-$1.5 $\pm$ ~1.5 \\ 
\enddata
\tablecomments{The Galactocentric coordinate ($R_s$, $\beta$) and the peculiar motion ($U_s$, $V_s$, $W_s$) of each source relative to the Galactic rotation model { A5} and solar motion from \cite{Reid:13}, described in the text. } 
 \label{tab:sctmot}
\end{deluxetable}

We now compare our result for the Scutum arm sources with HMSFRs in other spiral arms.
For 73 HMSFRs in the Sagittarius, Local, Perseus and Outer spiral arms (as listed in Table~1 of \citealt{Reid:13}),
 we obtain 
 { $\bar{U_s} = 3.7 \pm 0.7$~\kms, $\bar{V_s}= -1.5 \pm 1.1$~\kms, and $\bar{W_s}=0.6 \pm 0.6$~\kms\ }
 from the same model.
The Scutum arm sources show non-circular motions in the same direction as HMSFRs in other spiral arms
(\ie the same signs of $\bar{U_s}$ and $\bar{V_s}$), however, 
{ the motion toward the Galactic center is greater} for the Scutum arm sources.  
{ There are a few possible causes of the peculiar motions, as we discuss below.}
Firstly, the large non-circular motions may be due to the proximity of the arm { to} the gravitational potential
 of the central bar(s), as also observed for other HMSFRs closer to the Galactic center region
 \citep{Brunthaler:09, Sanna:09}.
{ The bar potential may give rise to highly non-circular stellar orbits or gas streamlines, such as 
 the $x_1, x_2$ and $x_3$ families of orbits \citep{Contopoulos:80, vanAlbada:82, Athanassoula:92}. 
However, our results contain no large peculiar velocities (beyond 30 \kms) that are significant at the 
 95\% level, which means the peculiar velocities are comparable to those in other arms at larger 
 Galactocentric radii. 
Hence large effects from the bar on our velocity measurements seem unlikely.}
{ A further potential origin for the peculiar velocities is the possibility that the Scutum arm itself is a mass
 overdensity \citep{Drimmel:01}, which may alter the 3-D kinematics itself. 
The uniformity in the signs of the peculiar velocities of the Scutum arm and other spiral arms may support
 this interpretation.}
 
{ The observed peculiar velocities could also be caused by an incorrect Solar peculiar motion,
 partially due to non-circular velocities of the local standard of rest (LSR).
\citet{Bovy:12} recently suggested the Solar rotational velocity ($V_\odot$) to be 24 \kms\ higher than 
 the circular rotation of the Galaxy of $\Theta_0 = 218 \pm 6$~\kms. 
While $\Theta_0$ and $V_\odot$ are strongly correlated, the sum of the two parameters (\ie the full tangential speed)
 is well constrained.
\citet{Bovy:12} find $\Theta_0 + V_\odot = 242^{+10}_{\,-3}$~\kms\
 ($V_\odot = 23.9^{+5.1}_{\,-0.5}$~\kms) and
 \cite{Reid:13} find $\Theta_0 + V_\odot = 255.2 \pm 5.1$~\kms.
As a test, if we assume $\Theta_0 = 222$~\kms, $V_\odot = 28$~\kms\ 
 (which gives $\Theta_0 + V_\odot = 250$~\kms) and
 a flat rotation curve and adopt $U_\odot = 10 \pm 1$~\kms\ \citep{Bovy:12} and $R_0=8.34\pm 0.16$~kpc \citep{Reid:13},
 we obtain $\bar{U_s} = 7.9 \pm 2.2$~\kms\ and $\bar{V_s}= 8.1 \pm 1.7$~\kms\  for the Scutum arm sources. 
(For comparison, for the 73 HMSFRs in other arms described above, 
 we obtain $\bar{U_s} = 3.7 \pm 0.6$~\kms\ and $\bar{V_s}= 11.5 \pm 0.6$~\kms.) 
Therefore, the rotational component ($\bar{V_s}$) of the peculiar motions of the Scutum arm sources 
 can be explained by an inaccurate value ($V_\odot$) for the Solar motion, but not the radial component ($\bar{U_s}$).  
Since the Scutum arm is located close to the Galactic center, where the rotation curve may begin to
 fall, this result is also noteworthy in terms of a slower circular rotation of the inner Galaxy (\eg \citealt{Honma:97}) and
 strengthens the result from our analysis that the velocity component $\bar{U_s}$
 toward the Galactic center is larger for the Scutum arm sources
 than for sources at larger Galactocentric radii.  
}


\acknowledgements
MS acknowledges financial support from a JSPS Postdoctoral Fellowship for Research Abroad.
This work was partially funded by the ERC Advanced Investigator Grant GLOSTAR (247078).
We thank the anonymous referee for insightful comments.

{\it Facilities:} \facility{VLBA}




\appendix
\twocolumn

\section{Description of the Target Masers}\label{app:description}
The general information of the observed sources can be found in Table~\ref{tab:source}.
A complete list of the observing dates and times is given in Table~\ref{tab:obs}.
Full details of the results from our parallax fits for the various maser spots with respect to the 
various reference quasars are given in Table~\ref{tab:parallaxes}. 
A discussion of historical work on these sources, along with a comparison with previous measurements, 
is given below.

\subsection{G005.88$-$00.39} 

G005.88$-$00.39, also known as W28 A2 (located $\sim 50'$ south of the W28 supernova remnant),
 is a well-studied shell-type ultracompact (UC) \hii region (\eg \citealt{Wood:89, Hunter:08})
 that consists of a shell of ionized gas surrounding a cavity.
A candidate ionizing star of spectral type O5 or earlier 
 has been detected in the near-infrared by \citet{Feldt:03}.

We measure a trigonometric parallax of $0.334\pm0.020$~mas for G005.88$-$00.39,
 corresponding to a distance of 2.99$^{+0.19}_{\,-0.17}$ kpc. 
This is consistent with several kinematic distance measures, \eg
 by  \citet[2.0$^{+0.7}_{\,-0.4}$ kpc using nebula expansion velocity]{Acord:98}
 and \citet[3.8$^{+1.0}_{\,-1.7}$ kpc]{Fish:03},
 but in disagreement with the kinematic distance to the supernova remnant W28
 measured by \citet[$1.9\pm 0.3$~kpc]{Velazquez:02}.
Our measurement is significantly inconsistent with
 the recent trigonometric parallax measurement by \cite{Motogi:11}
 of $0.78\pm0.05$~mas (distance of 1.28$^{+0.09}_{\,-0.08}$ kpc) using VERA. 
They traced 2 \hho maser feature positions in right ascension (R.A.) of G005.88$-$00.39 at 8 epochs.
From the standard deviations of the parallax fits to the two features, which lie between 0.3 to 0.4 mas, 
 they derived a least-squares formal fitting error of $\sim$0.04 mas in the measured parallax.
They discuss potential additional astrometric errors introduced by atmospheric zenith delay residuals
 and the quality of the phase-referenced map to be $\sim$0.09 mas
  (this would increase by $\sqrt{2}$ without their assumption of no correlation between two maser feature positions),
 however, they conclude that their measurement error is dominantly caused by the intrinsic maser structure,
 which is reduced by $\sqrt{7}$ for having seven maser spots each in the two features.
Since they used only one background source, there is room
 for an unseen contribution to their measurement errors, dominantly caused by the atmospheric zenith delay residuals.
In comparing our own result with those of \cite{Motogi:11},
 we consider our own measurements more reliable, since we used two different background sources,
 thereby limiting potential position errors caused by the residuals of the atmospheric zenith delays.
Moreover, we are advocating a more conservative error estimate as mentioned above.

\begin{figure*}
	\centering
   \includegraphics[width=8cm, angle=270]{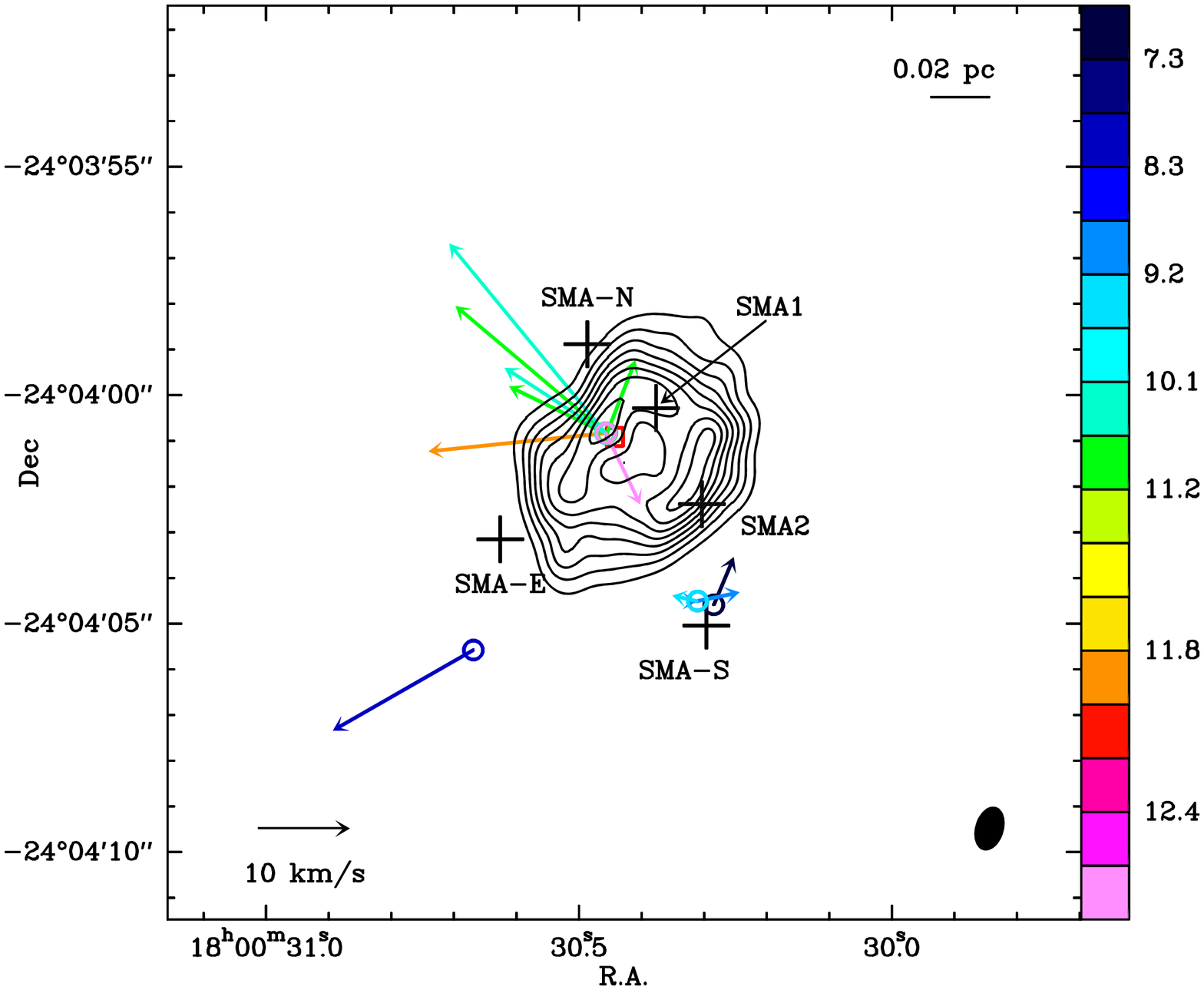}
		\caption{\footnotesize Relative proper motions of the detected maser spots in G005.88$-$00.39 with respect to the reference spot at \vlsr= 9.4 \kms, overlaid with a continuum map at 8.4 GHz from VLA archive data.}\label{fig:G005spotmap}
\end{figure*}

Figure~\ref{fig:G005spotmap} shows the relative proper motions 
 of the detected maser spots in G005.88$-$00.39 with respect to the reference spot at \vlsr = 9.4 \kms, 
 overlaid with a continuum map at 8.4 GHz from VLA archive data.
The position of the ionizing star, suggested by \cite{Feldt:03}, is marked by a red open square with the size of the position error. 
Crosses indicate the postions of five dust emission sources observed at 875 $\mu$m with the Submillimeter Array (SMA) by \cite{Hunter:08}.

\subsection{G011.91$-$00.61}

The source G011.91$-$00.61 
 is associated with an infrared dark cloud \citep{Cyganowski:08}. 
It was identified as an ``extended green object'' (EGO) by \citet{Cyganowski:08}
in images of the Spitzer/GLIMPSE (Galactic Legacy Infrared Mid-Plane Survey Extraordinaire)
surveys \citep{Churchwell:09} showing strong emission at 4.5 $\mu$m band images
made with the InfraRed Array Camera (IRAC),
which is believed to trace shocked gas in protostellar outflows. 
Besides water maser emission \citep[detected by][]{Hofner:96},
weak emission of 6.7-GHz methanol masers was detected in two kinematically distinct groups,
separated by $\sim$5$\arcsec$ \citep{Cyganowski:09}. 
The maser groups coincide with the submillimeter clumps MM1 and MM3,
observed at 1.3 mm \citep{Cyganowski:11}. 
The velocities of the methanol masers cover a range from 25 to 57 km s$^{-1}$
and are redshifted by a few km s$^{-1}$ from the thermal gas velocity V$_{ \rm LSR}$
\citep[36 km s$^{-1}$ from H$^{13}$CO$^{+}$ observations,][]{Cyganowski:09}. 
From the systemic velocity of the thermal gas, \citet{Cyganowski:09} determined
 a near-kinematic distance to G011.91$-$00.61 of 3.8$^{+ 0.40}_{- 0.46}$ kpc.
The submillimeter spectrum of MM1 at 230 GHz shows rich molecular line emission,
 similar to the spectra of hot cores. 
In MM3, compact emission arises only from C$^{18}$O \citep{Cyganowski:11}. 
Channel maps of $^{12}$CO(2-1) show evidence for multiple outflows, centered on MM1,
 which are probably driven by a still accreting massive young stellar object \citep{Cyganowski:11}. 

We observed G011.91$-$00.61 at nine epochs 
from 2010 September to 2012 January (Table~\ref{tab:obs}). 
However, due to bad weather conditions, the data of the first epoch could not be used for the parallax measurements. 
We observed three different quasars (J1808--1822, J1809--1520, and J1825--1718; Table~\ref{tab:source}) as background position references. 
Since the quasar J1809--1520 was detected in only four and not in all epochs, 
 we did not use this source for the parallax determination.
We used a circular restoring beam of 2 mas (HPBW) for the maser and quasar images of all epochs.

We measure a trigonometric parallax of $0.297\pm0.031$~mas for G011.91$-$00.61,
 corresponding to a distance of 3.37$^{+0.39}_{\,-0.32}$ kpc (see Fig. \ref{fig:parallaxes}b). 
Our result is consistent within the errors with the near-kinematic distance.

\begin{figure*}
	\centering
   \subfigure{\includegraphics[width=8cm]{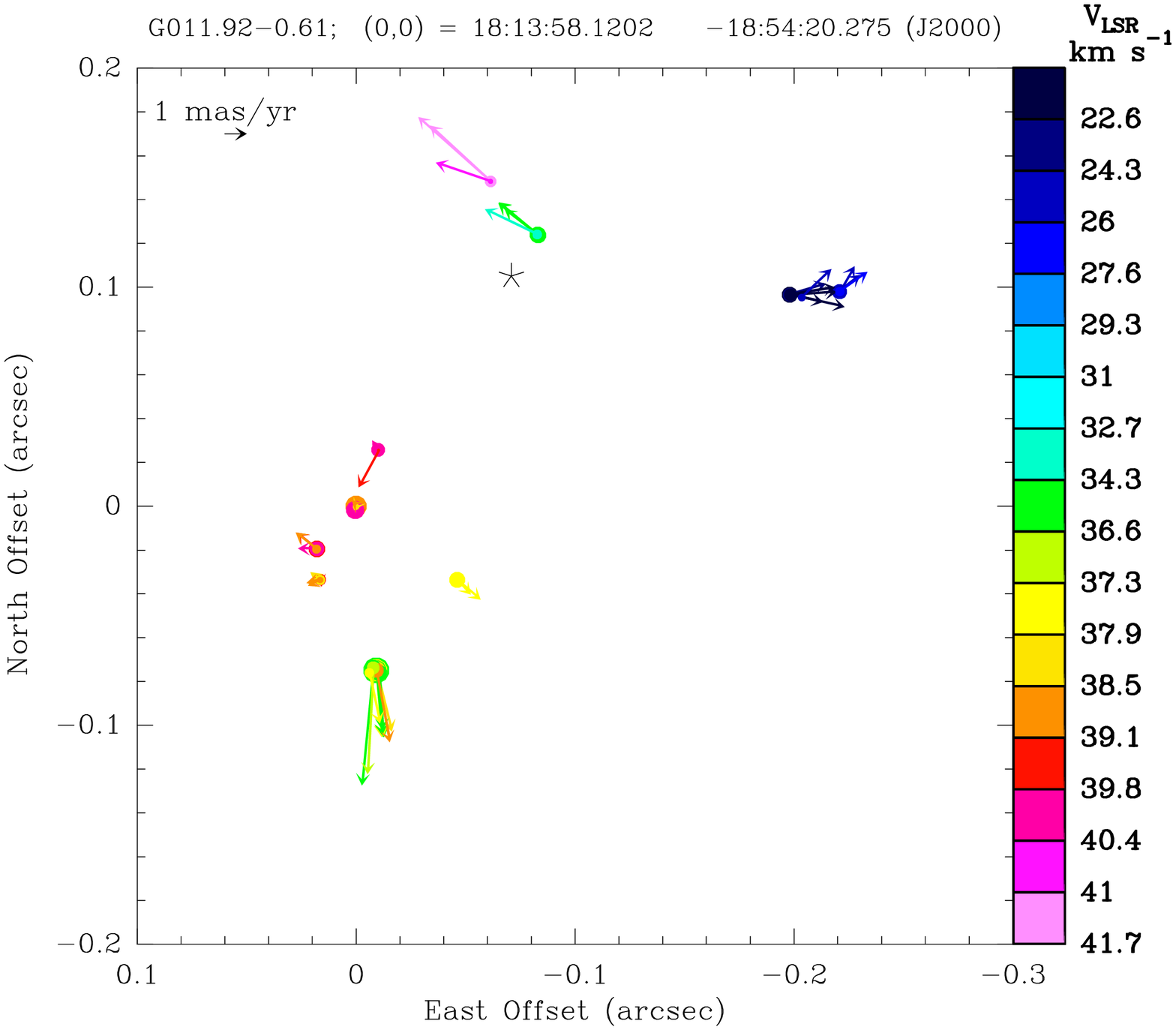}\label{fig:G011spotmap}}
   \subfigure{\includegraphics[width=8cm]{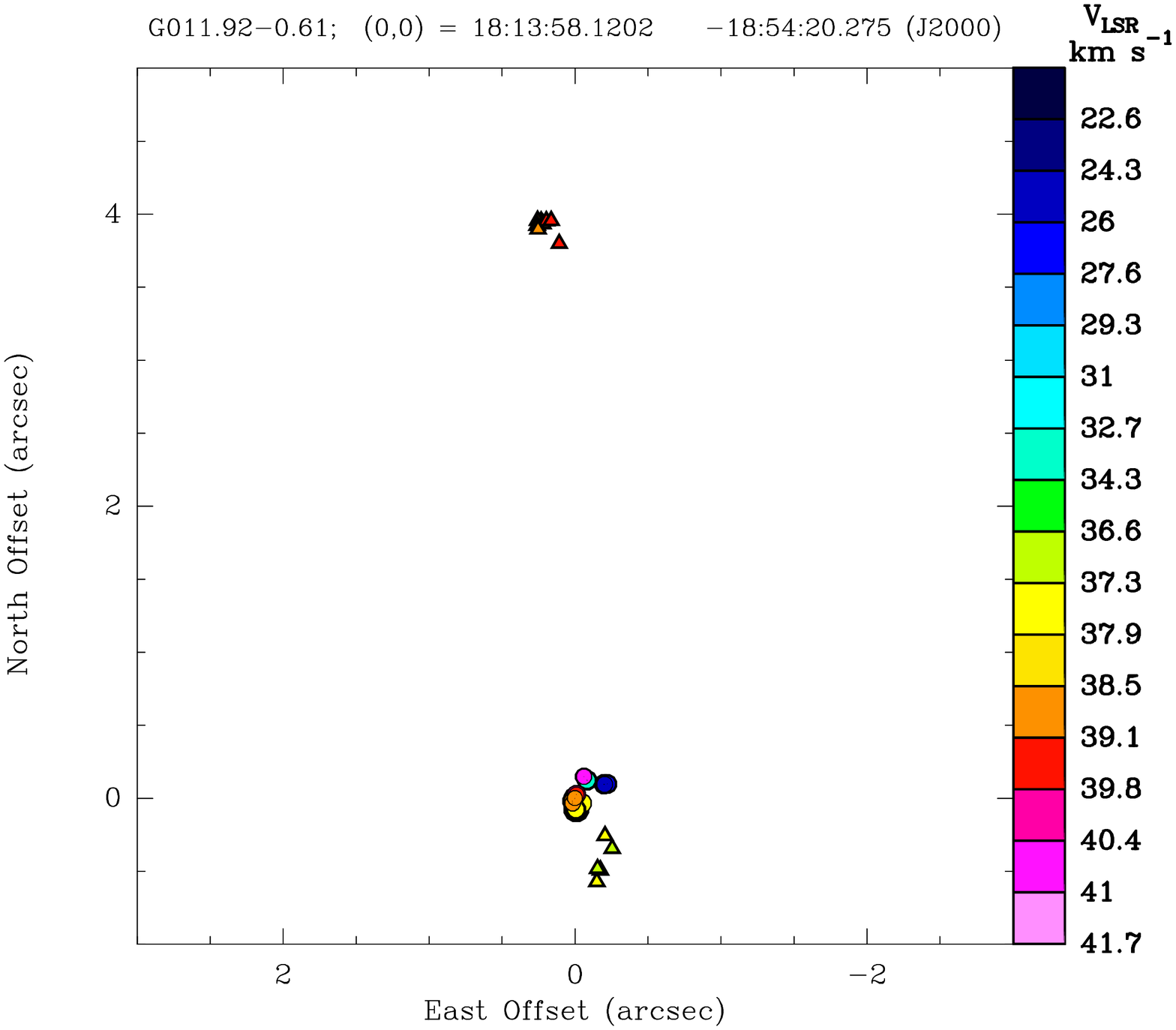}\label{fig:G011spotmapMethMaser}}
		\caption{\footnotesize Left panel: Positions of all maser spots in G011.91--0.61 that have been detected in at least three epochs (positions from epoch 2011 May 21). The size of the spots scales linearly with the flux density of the spots, \ie larger spots correspond to higher flux densities. The arrows show the proper motions of the maser spots in the reference frame of the central star, marked by an asterisk. The length of the arrows is enhanced by a factor of 10 for a better visibility. Right panel: Water maser spot positions (marked by circles) relative to the positions of the 6.7-GHz methanol masers (marked by triangles), detected by \citet{Cyganowski:09}.}
	\label{fig:spotmapG011}
\end{figure*}

In our observations of G011.91$-$00.61, we detected several strong water maser spots in an area of 0.5$\arcsec$~x~0.5$\arcsec$, covering a velocity range of 20 to 42 km s$^{-1}$. 
Figure~\ref{fig:spotmapG011} shows the maser spot map with the proper motions of each maser spot. 
The maser spots can be separated into two different groups, probably tracing two outflows almost perpendicular to each other. 
Their positions and angles are roughly consistent with the large scale bipolar outflows, detected in $^{12}$CO observations \citep{Cyganowski:11}. 
We fitted the proper motions of the maser spots with a model of uniformly expanding flows \citep{Sato:10W}, yielding the position and proper motions of the central driving source
 (marked by an asterisk in Figure~\ref{fig:G011spotmap}). 
The fitted absolute proper motions of the driving source are $\mu_{x}$ = (0.66 $\pm$ 0.28) mas yr$^{-1}$ eastward and $\mu_{y}$ = ($-$1.36 $\pm$ 0.41)  mas yr$^{-1}$ northward.

Figure~\ref{fig:G011spotmapMethMaser} shows the positions of the water maser spots
 relative to the positions of the 6.7-GHz methanol masers, detected by \citet{Cyganowski:09}. 
The water maser spots are located close to the southern group of methanol masers and
 are thus also located in the submillimeter clump MM1. 
No water maser emission was detected around the positions of the northern methanol maser group. 
The line-poor spectrum of MM3 and the non-detection of water masers indicate that this clump is in
 an earlier stage of evolution than MM1 with the only sign for ongoing star formation being the detection of 6.7-GHz methanol masers.

\subsection{G013.87$+$00.28} 

The 22-GHz \hho maser was first detected by \cite{Jaffe:81} from the sample of far-infrared sources.
We observed three background sources (J1809$-$1520, J1808$-$1822 and J1825$-$1718; see Table~\ref{tab:source}), but 
 we used only the closest source J1809$-$1520 to the maser target on the sky for parallax measurement.
This source showed a double lobe structure with the two lobes clearly separated at each epoch, which was stable in time.
J1808$-$1822 was not detected in the middle epochs and thus was not used for the parallax fitting. 
J1825$-$1718, the most distant background source from the maser on the sky, showed structural change 
 likely due to the phase correction residuals from the maser target.
The background sources were convolved with a circular restoring beam of 0.4~mas (HPBW) at each
 epoch and uniform weighting.  
 
Using six maser spots in two maser features, we measure a parallax of $0.254\pm0.024$~mas,
 corresponding to a distance of 3.94$^{+0.41}_{\,-0.34}$ kpc, which is consistent with but 
 slightly closer than the kinematic distances of $4.6\pm0.5$ kpc by \cite{Sweilo:04} and 4.5 kpc by \cite{Thompson:06}.

\subsection{G016.58$-$00.05} \label{sect:g16}

G016.58$-$00.05 (IRAS 18182$-$1433) is
 considered to be a high-mass protostellar object 
 at an early evolutionary stage prior to an  
 UC\hii region \citep{Sridharan:02}.
 \cite{Beuther:06} found multiple massive outflows in the region from the SMA observations, 
 indicative of mutiple young stellar objects across a region of $\approx0.1$ pc \citep{Sanna:10} .

We observed three background sources (J1825$-$1551, J1825$-$1718, and J1809$-$1520; Table~\ref{tab:source}).
Since the angular separation on the sky between the background and maser sources is 
 much smaller for J1825$-$1551 (1\d7 separation) than for the other two sources (3\d0 separation),
 the parallax fitting errors were significantly smaller using J1825$-$1551 than using one of the other two sources (Table~\ref{tab:parallaxes}).
Therefore, separate error floors were used for each background source when doing independent fits,
 and for the combined fit these floors were adopted.
   
Using five maser spots in two features, we measure a parallax of $0.279\pm0.023$ mas for G016.58$-$00.05, corresponding to a distance of 3.58$^{+0.32}_{\,-0.27}$ kpc.
This is substantially closer than the near kinematic distance by \citet[5.4 kpc]{Walsh:97}.

\begin{figure}
	\centering
   \includegraphics[width=7cm, angle=270]{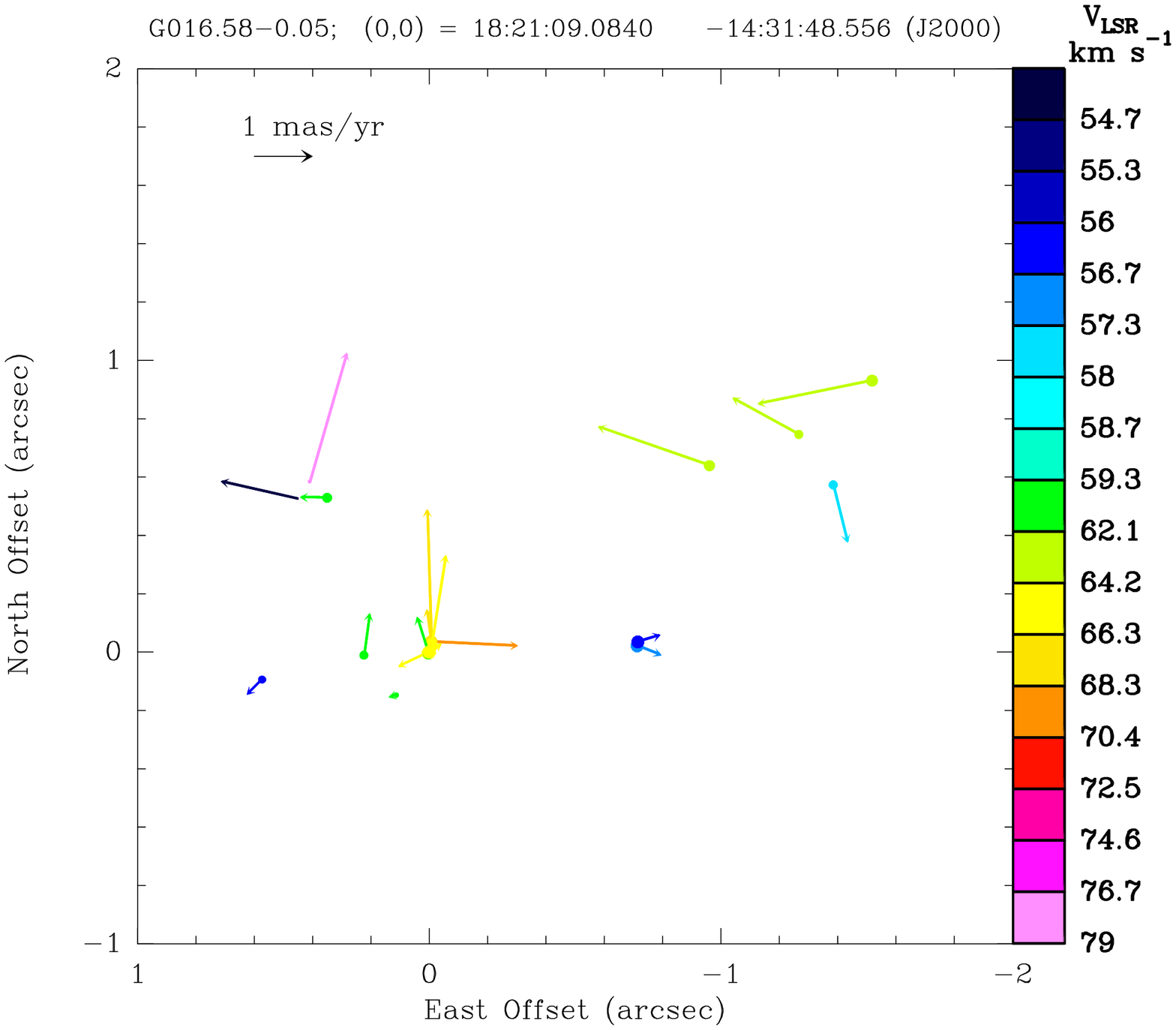}\label{fig:G016spotmap}
		\caption{\footnotesize Positions and motions of all maser features in G016.58$-$00.05 that have been detected in at least three epochs (positions are fitted for the mean epoch, year 2011.696).
		The size of the spots scales linearly with the flux density of the spots, \ie larger spots correspond to higher flux densities. 
		The arrows show the proper motions of the maser spots relative to the reference spot at $\vlsr = 64.3$~\kms\ located at the map origin.
		The length of the arrows is enhanced by a factor of 200 for a better visibility.}
\end{figure}

Based on the criteria as in \cite{Sato:10W}, we detected total 21 maser features in at least 3 epochs over all 11 epochs. 
Figure~\ref{fig:G016spotmap} shows the position and proper motions of each maser feature relative to the reference maser spot at $\vlsr = 64.3$~\kms\ located at the map origin. 
The proper motions and large spatial distribution ($\sim 3'' \sim 0.78$~pc at a distance of 3.58 kpc) indicate that the maser features are likely driven by at least two young stellar objects.
In order to estimate the systemic motion of the region, we adopted an unweighted mean of the proper motions of all detected 21 maser features, 
 $\mu_{x}$ = ($0.20 \pm 0.18$) mas yr$^{-1}$ eastward and $\mu_{y}$ = (0.36 $\pm$ 0.17)  mas yr$^{-1}$ northward with respect to the reference spot (at $\vlsr = 64.3$~\kms). 
Adding the absolute proper motion of the reference spot from parallax measurement and including additional 5\kms\ uncertainty in each direction, we obtain 
 the systemic absolute proper motions to be $\mu_{x}$ = ($-1.13 \pm 0.34$) mas yr$^{-1}$ eastward and $\mu_{y}$ = ($-$2.59 $\pm$ 0.35)  mas yr$^{-1}$ northward.

\subsection{G025.70$+$00.04} 
\label{subsec:g25} 

\begin{figure}[H]
  \centering \includegraphics[angle=-90,scale=0.42]{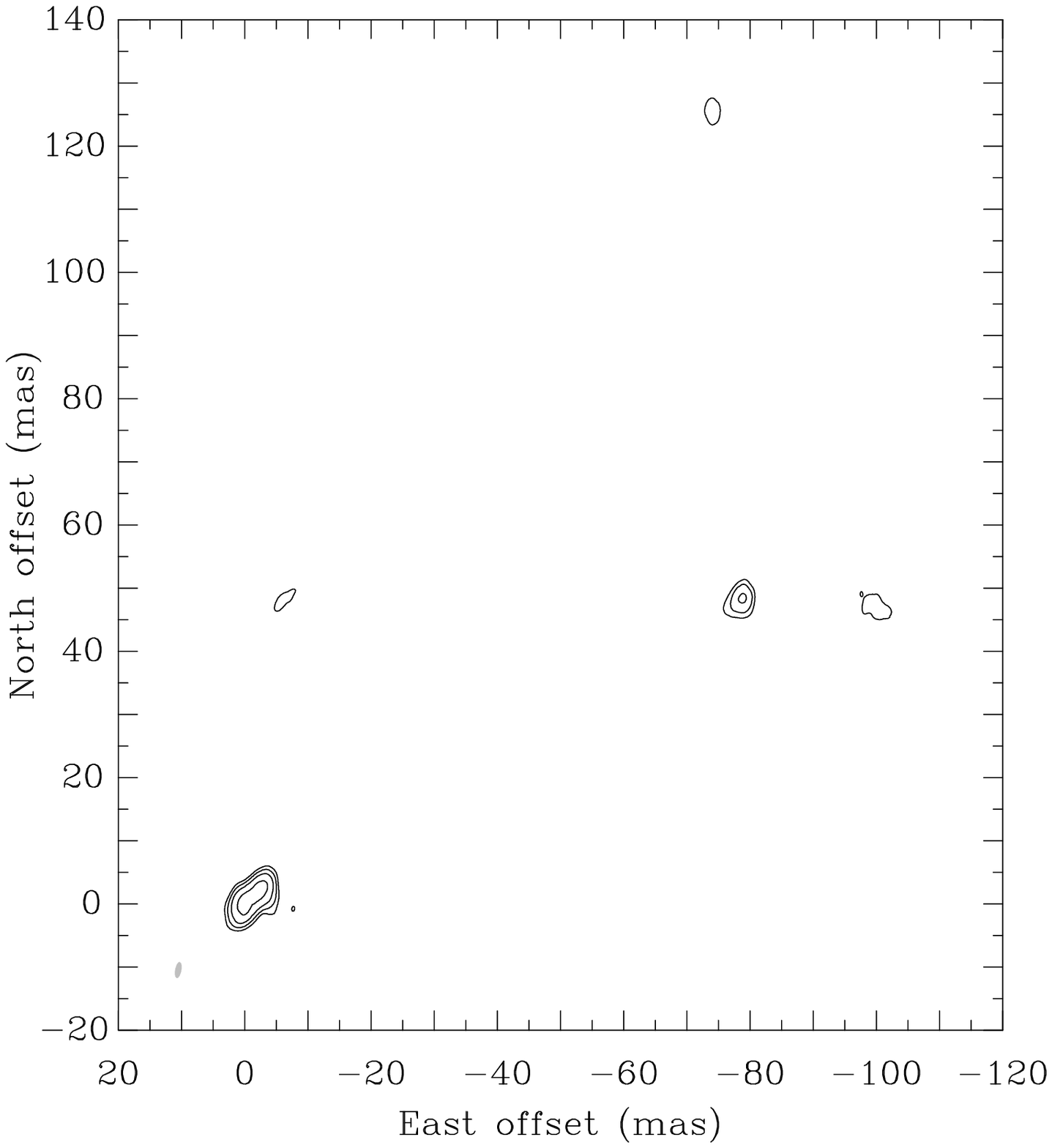}
  \caption{\footnotesize Images of velocity-integrated maser emission of G025.70$+$00.04 at the first epoch.  
Contour levels start at 450 Jy~beam$^{-1}$~km~s$^{-1}$ and
 increase linearly.  Restoring beam (gray) is in the lower left corner. }
  \label{fig:g25_mom0}
\end{figure}

The 6.7-GHz methanol maser in G025.70$+$00.04 was discovered from IRAS (the Infrared Astronomical Satellite) selected
 candidates by \citet{Walsh:97}.  From this sample, the 12-GHz methanol maser
 was first detected by \citet{Blaszkiewicz:04}.  
This star formation region also contains OH and water masers \citep{Szymczak:04,Szymczak:05}.  
The 12-GHz methanol maser emission toward G025.70$+$00.04 consists of maser spots separated by
 $\sim$ 140 mas (see Fig.~\ref{fig:g25_mom0}).  
We detected three maser spots at \vlsr\ = 95.4, 94.6, and 89.6 \kms\ that were suitable for parallax fitting.  
These three maser spots are located at offsets of about (0, 0), ($-$70, 125) and ($-$75, 50) mas
 in Fig.~\ref{fig:g25_mom0}, respectively.  
The background source J1837$-$0653 and J1846$-$0651 were detected at all epochs. 
We found that only phase referencing to J1837$-$0653, which is much closer in angular distance to the target source G025.70$+$00.04
 than J1846$-$0651, can give reasonable parallaxes.  

The results of parallax fitting are given in Table~\ref{tab:parallaxes}.  
We adopt the ``combined'' parallax solution, which uses three maser spots and one
 background source.  
We do not use the
 formal parallax uncertainty of $\pm0.017$~mas, but instead conservatively
 multiply the formal uncertainty by $\sqrt{3}$.  
Thus, we obtain a parallax for G025.70$+$00.04 of $0.098 \pm 0.029$~mas,
 corresponding to a distance of $10.20^{+4.38}_{\,-2.38}$~kpc.  
The trigonometric distance is consistent with the far kinematic distance
 of 9.9 kpc with averaged $\vlsr = 93.2$ \kms, using IAU Galactic values,
 rather than the near kinematic distance of 5.4~kpc.  

The proper motion estimates for the three maser spots appear consistent within
 their joint uncertainties and the average proper motion is $-2.89 \pm
 0.05$~\masy\ toward the east and $-6.21 \pm 0.35$~\masy\ toward the north. 
For the measured parallax, the proper motion corresponds to $-139.7$~\kms\ and
 $-300.2$~\kms\ eastward and northward, respectively. 
In Fig.~\ref{fig:parallaxes}(f), we plot the positions of these maser spots relative to
 the background radio source with superposed curves representing the model maser
 tracks across the sky.

\subsection{G032.04$+$00.05} 

The 6.7-GHz methanol maser in G032.04$+$00.05 was first detected in color-selected IRAS sources
 of ultra-compact \hii regions by \cite{vanderWalt:95}.
The 12.2-GHz maser in this source was detected by \cite{Gaylard:94}
 and was confirmed to be in the same emitting region as the 6.7-GHz methanol maser
 by \cite{Blaszkiewicz:04}.
\cite{Ellingsen:11} also detected methanol maser emission at 37.7 and 38.3 GHz in this source,
 which are two other known class\ion{}{2} methanol maser transitions \citep{Haschick:89}.
\cite{Ellingsen:11} suggest that the 37.7-GHz maser emission is associated with
 a brief evolutionary stage $1000-4000$ years prior to the cessation of class\ion{}{2} maser activity.

We observed the 12-GHz methanol maser in G032.04$+$00.05
 with four background sources (Table~\ref{tab:obs}).
The background source J1857$-$0048 was closer to the target maser source on the sky (2\d1 separation) by a factor of two 
 than other background sources, and thus yielded the position measurement
 least affected by the residuals of atmospheric zenith delay calibration.
Using J1857$-$0048, we obtained a parallax of 0.209$\pm0.011$~mas.
We excluded the furthest background source on the sky, J1907$+$0127 (4\d9 separation), from the combined parallax measurement,
 since it yielded clearly worse fit results (Table ~\ref{tab:parallaxes}) than with other background sources due to the atmospheric zenith delay residuals.
The other two background sources, J1904$+$0110 and J1833$-$0323, are located on the sky on opposite sides of the target maser, 
 and thus the average from these two sources is expected to yield a parallax less affected from the atmospheric delay residuals than the individual fits,
 assuming the atmospheric perturbation is linear.
The average value of the parallaxes from the two sources is 0.1825 mas.

Considering the $\approx$4$^\circ$ separation of each source from the target maser source, which is a factor of two larger than that of J1857$-$0048,
 we weighted these two sources down by a factor of two in calculating the combined parallax with J1857$-$0048.
Therefore, the final parallax was calculated as a weighted mean of the individual fits using the three background sources with weights of 2:1:1 for 
 J1857$-$0048, J1904$+$0110, and J1833$-$0323.
For the uncertainty estimate, we treated J1857$-$0048 and the average from  J1904$+$0110 and J1833$-$0323
 as we would for two background sources. 
The parallax uncertainty using a single background source J1857$-$0048, 0.011~mas, 
 divided by $\sqrt{2}$ (for two background sources) yields an uncertainty of 0.008~mas.
Also, half the difference of the two parallaxes (0.209 mas and 0.1825 mas) yields an uncertainty of 0.008~mas.
Therefore, adopting 0.008 mas as the parallax uncertainty, we obtain a final parallax of $0.193\pm0.008$~mas.
This corresponds to a distance of $5.18^{+0.22}_{\,-0.21}$~kpc.

\setcounter{table}{0}
\renewcommand\thetable{\Alph{section}.\arabic{table}}

\begin{deluxetable}{lllrrllll}
\tabletypesize{\scriptsize}
\tablecaption{Source information \label{tab:source}}
\tablewidth{0pt}
\tablehead{
\colhead{Source} & \colhead{R.A.(J2000)} & \colhead{Dec.(J2000)} &
\colhead{$\theta_{\rm sep}$}  & \colhead{P.A.} & \colhead{Restoring Beam} & 
\colhead{$F_{\rm peak}$} & \colhead{$V_{\rm LSR}$} &\colhead{Ref.}\\
  & ~~~~(h m s) & ~~~~~~~($^{\circ}$ $^{\prime}$ $^{\prime\prime}$) & ($^{\circ}$) &
($^{\circ}$) &  (mas $\times$ mas at $^{\circ}$) & (Jy beam$^{-1}$) & (km s$^{-1}$)
}
\startdata
G005.88$-$00.39 & 18 00 30.3061  & $-$24 04 04.478   & ...    & ...        & 0.93$\times$0.30 at $-$2 & 9.6       & ~11.9 &\\
J1807$-$2506     & 18 07 40.6877   & $-$25 06 25.943   & 1.9 & 122     & 1.32$\times$0.34 at $+$1 &  ~0.08 & ~~... & 1 \\ 
J1755$-$2232     & 17 55 26.2848   & $-$22 32 10.617   & 1.9 & $-$37 & 1.25$\times$0.33 at $-$2 & ~0.15 & ~~... & 1 \\ 
J1751$-$1950     & 17 51 41.3438   & $-$19 50 47.506   & 4.7 & $-$25 & 1.15$\times$0.34 at $-$4 & ~0.02 & ~~... & 2 \\ 
\hline
G011.91$-$00.61 & 18 13 58.1170   & $-$18 54 20.353    & ...     & ...       & 2.0$\times$2.0 at ...   & 12.3~ & ~39.4 & \\ 
J1808$-$1822    & 18 08 55.51496 & $-$18 22 53.3832 & 1.3   & $-$66 & 2.0$\times$2.0 at ...  & ~0.02 & ~~... & 2 \\
J1825$-$1718    & 18 25 36.5323   & $-$17 18 49.848   & 3.2   &       60 & 2.0$\times$2.0 at ...  & ~0.09 & ~~... & 1 \\
J1809$-$1520    & 18 09 10.20936 & $-$15 20 09.6991 & 3.7  & $-$18 & 2.0$\times$2.0 at ...  & ~0.01 & ~~... & 2 \\
\hline
G013.87$+$00.28 & 18 14 35.8324    & $-$16 45 35.867    & ...   & ...         & 0.4$\times$0.4 at ... & 16.6 & $-$26.8 & \\ 
J1809$-$1520     &  18 09 10.20936 & $-$15 20 09.6991 & 1.9 &   $-$42 & 0.4$\times$0.4 at ... & ~0.01 & ~~... & 2 \\
J1808$-$1822     &  18 08 55.51535 & $-$18 22 53.3961 & 2.1 & $-$140 & 0.4$\times$0.4 at ... & ~0.01 & ~~... & 2 \\ 
J1825$-$1718     &  18 25 36.5323   & $-$17 18 49.848    & 2.7 &       102 & 0.4$\times$0.4 at ... & ~0.03 & ~~... & 1 \\ 
\hline
G016.58$-$00.05 & 18 21 09.0808   & $-$14 31 48.784    & ...   &  ...         & 0.88$\times$0.33 at $-$14 & 9.7~ & ~64.2 & \\ 
J1825$-$1551    & 18 25 11.72184 & $-$15 51 34.4066 & 1.7 &       144 & 1.15$\times$1.08 at $-$13  &  ~0.14 & ~~... & 2 \\
J1825$-$1718    & 18 25 36.5323   & $-$17 18 49.848    & 3.0 &       159 & 1.20$\times$1.06 at $-$44 & ~0.57 & ~~... & 1 \\
J1809$-$1520    & 18 09 10.20936 & $-$15 20 09.6991 & 3.0 & $-$106 & 1.64$\times$0.93 at $-$88 & ~0.16 & ~~... & 2 \\ 
\hline
G025.70$+$00.04 & 18 38 03.145        & $-$06 24 15.19        & ...    & ...     &  2.1$\times$1.3 at $-$4  & 10.0~ & ~95.4 &  \\
J1837$-$0653     & 18 37 58.039        & $-$06 53 31.00        & 0.5 & $-$178 & 2.3$\times$1.3 at 0 & ~0.01 & ~~... & 3 \\ 
J1846$-$0651     & 18 46 06.300263 & $-$06 51 27.74561 & 2.1 &       103 & 2.3$\times$1.5 at $+$1 & ~0.03 & ~~... & 4 \\ 
\hline
G032.04$+$00.05 & 18 49 36.5761    & $-$00 45 46.899    & ... & ...           & 1.65$\times$0.69 at  $+$3 &  ~4.2~ & ~98.8 & \\ 
J1857$-$0048     & 18 57 51.35860  & $-$00 48 21.9496 & 2.1 &         91 & 2.07$\times$0.86 at  $+$14 & ~0.05 & ~~... & 2 \\
J1904$+$0110    & 19 04 26.3975    & $+$01 10 36.703   & 4.2 &         62 & 1.99$\times$0.82 at $+$15 & ~0.06 & ~~... & 2 \\ 
J1833$-$0323     & 18 33 23.90553  & $-$03 23 31.4656 & 4.8 & $-$123 & 2.10$\times$0.84 at $+$11 & ~0.05 & ~~... & 2 \\
J1907$+$0127    & 19 07 11.9962    & $+$01 27 08.964   & 4.9 &         63 & 2.06$\times$0.85 at $+$15 & ~0.14 & ~~... & 1\\
\enddata
\tablecomments{Positions and source properties for the target maser and the background quasar calibrators.
The first column gives the source names.  
Columns 2 and 3 list the observed source positions,
Cols.\ 4 and 5 give the angular separation and position angles (East of North) of background calibrators
  relative to the maser targets, respectively.   
Column 6 lists the interferometer restoring beam (FWHM). 
Column 7 lists the peak flux densities and Column 8 lists the radial velocities (with respect to the Local Standard of Rest)
 of the reference maser spots, obtained at epoch 2012 Feb 11 for G005.88$-$00.39,
 at epoch 2011 May 21 for G011.91$-$00.61,
 at epoch 2012 Sep 23 for G016.58$-$00.05,
 and at the first epoch for the other sources.}
\tablerefs{
(1) \cite{Ma:09}
(2) \cite{Immer:11}
(3) \cite{Xu:09}
(4) \cite{Petrov:05}
}
\end{deluxetable}

\begin{deluxetable}{lccclll}
\tabletypesize{\scriptsize}
\tablecolumns{6} \tablewidth{0pc} 
\tablecaption{VLBA Observations}
\tablehead {
  \colhead{Source} & \colhead{Line} & \colhead{Project} &
  \colhead{Epoch} & \colhead{Date} & 
  \colhead{Time Range} 
\\
  \colhead{}  &  \colhead{} & \colhead{}  & \colhead{}     & \colhead{} & 
  \colhead{(UT)}     
            }
\startdata
G005.88$-$00.39 & 22 GHz & BR145R  & 1 & 2010 Sep 10 & 22:03--04:58\tablenotemark{a} \\
                  &                               &                & 2 & 2010 Dec 14 & 15:50--22:44  \\
                  &                               &                & 3 & 2011 Feb 11 & 11:58--19:01  \\
                  &                               &                & 4 & 2011 Mar 18 & 09:40--16:43 \\
                  &                               &                & 5 & 2011 May 12 & 06:32--12:32 \\
                  &                               &                & 6 & 2011 Sep 11 & 22:00--05:03\tablenotemark{a} \\
                  &                               &                & 7 & 2011 Dec 16 & 15:43--22:46 \\
                  &                               &                & 8 & 2012 Feb 11 & 11:59--19:02 \\
\hline
G011.91$-$00.61  & 22 GHz & BR145I & 1 & 2010 Sep 19 & 22:08--04:57\tablenotemark{a} \\
                &                                &               & 2 & 2010 Oct 22 & 19:58--02:47\tablenotemark{a} \\
                &                                &               & 3 & 2010 Dec 19 & 16:10--22:59\tablenotemark{a} \\
                &                                &               & 4 & 2011 Feb 15 & 12:22--19:17 \\
                &                                &               & 5 & 2011 Mar 26 & 09:49--16:44 \\
                &                                &               & 6 & 2011 May 21 & 06:08--13:04 \\
                &                                &               & 7 & 2011 Jul 30  & 01:33--08:29 \\ 
                &                                 &              & 8 & 2011 Sep 20 & 22:05--05:00\tablenotemark{a} \\                
\hline
G013.87$+$00.28   & 22 GHz  & BR145M & 1 & 2011 Mar 19 & 10:16--17:11 \\
                    &                               &                & 2 & 2011 Jun 20 & 04:10--11:05 \\
                    &                               &                & 3 & 2011 Aug 20 & 00:10--07:05 \\
                    &                               &                & 4 & 2011 Oct 02 & 21:18--04:12\tablenotemark{a} \\ 
                    &                               &                & 5 & 2011 Nov 28 & 17:33--00:28\tablenotemark{a} \\
                    &                               &                & 6 & 2012 Mar 22 & 10:01--16:56 \\
\hline
G016.58$-$00.05    & 22 GHz  & BR145U & 1 & 2010 Sep 20 & 21:54--04:44\tablenotemark{a}\\
                     &                              &                 & 2 & 2010 Dec 23 & 15:44--22:34 \\
                     &                              &                 & 3 & 2011 Feb 16 & 12:08--19:06 \\
                     &                              &                 & 4 & 2011 Mar 28 & 09:31--16:29 \\
                     &                              &                 & 5 & 2011 May 25 & 05:43--12:41 \\
                     &                              &                 & 6 & 2011 Sep 25 & 21:35--04:33\tablenotemark{a}\\
                     &                              &                 & 7 & 2011 Dec 31 & 15:14--22:12 \\
                     &                              &                 & 8 & 2012 Feb 16 & 12:09--19:07 \\
                     &                              &                 & 9 & 2012 Mar 26 & 09:36--16:34 \\
                     &                              &               & 10 & 2012 May 13 & 06:27--13:25 \\
                     &                              &               & 11 & 2012 Sep 23 & 21:40--04:38\tablenotemark{a}\\
\hline
G025.70$+$00.04 & 12 GHz & BR129B & 1 & 2007 Oct 21 & 19:13--04:03\tablenotemark{a} \\
                                &                &                 & 2 & 2008 Apr 18 & 07:25--16:21 \\
                                &                &                 & 3 & 2008 Sep 21 & 21:08--06:04\tablenotemark{a} \\
                                &                &                 & 4 & 2009 Mar 28 & 08:49--17:45 \\
\hline
G032.04$+$00.05  & 12 GHz  & BR145Q &  1 & 2011 Mar 20 & 10:22--17:22 \\
                 &                              &                & 2 & 2011 Sep 15 & 22:34--05:34\tablenotemark{a} \\
                 &                              &                & 3 & 2011 Sep 26 & 21:51--04:51\tablenotemark{a} \\
                 &                              &                & 4 & 2012 Mar 24 & 10:03--17:03 \\
\enddata
\tablenotetext{a}{The observation ended on the day following the one listed.}
\label{tab:obs}
\end{deluxetable}

\begin{deluxetable}{lccccc}
\tabletypesize{\scriptsize}
\tablecaption{Detailed results of parallax and proper motion \label{tab:parallaxes}}
\tablewidth{0pt}
\tablehead{ 
\colhead{Maser Source} & \colhead{V$_{\rm LSR}$} & \colhead{Background} & \colhead{Parallax} &
\colhead{$\mu_{x}$}  & \colhead{$\mu_{y}$} \\
 & (km s$^{-1}$)  & Source & (mas) & (mas yr$^{-1}$) & (mas yr$^{-1}$)
}
\startdata
G005.88$-$00.39  & ~6.9 & J1807$-$2506 & 0.359 $\pm$ 0.027 & $-$0.54 $\pm$ 0.04 & $-$2.03 $\pm$ 0.10 \\
                                  & ~6.9 & J1755$-$2232 & 0.314 $\pm$ 0.036 & $-$0.52 $\pm$ 0.06 & $-$1.71 $\pm$ 0.17 \\
                                  & ~7.3 & J1807$-$2506 & 0.371 $\pm$ 0.027 & $-$0.57 $\pm$ 0.04 & $-$1.99 $\pm$ 0.09 \\
                                  & ~7.3 & J1755$-$2232 & 0.327 $\pm$ 0.031 & $-$0.55 $\pm$ 0.05 & $-$1.66 $\pm$ 0.17 \\
                                  & ~9.0 & J1807$-$2506 & 0.323 $\pm$ 0.032 & $-$0.37 $\pm$ 0.05 & $-$2.40 $\pm$ 0.11 \\
                                  & ~9.0 & J1755$-$2232 & 0.279 $\pm$ 0.021 & $-$0.35 $\pm$ 0.03 & $-$2.08 $\pm$ 0.13 \\
                                  & ~9.4 & J1807$-$2506 & 0.306 $\pm$ 0.019 & $-$0.56 $\pm$ 0.03 & $-$2.36 $\pm$ 0.11 \\
                                  & ~9.4 & J1755$-$2232 & 0.262 $\pm$ 0.031 & $-$0.54 $\pm$ 0.05 & $-$2.03 $\pm$ 0.13 \\
                                  & 10.3 & J1807$-$2506 & 0.383 $\pm$ 0.020 & $+$0.77 $\pm$ 0.03 & $-$1.33 $\pm$ 0.09 \\
                                  & 10.3 & J1755$-$2232 & 0.336 $\pm$ 0.027 & $+$0.78 $\pm$ 0.04 & $-$0.99 $\pm$ 0.16 \\
                                  & 10.7 & J1807$-$2506 & 0.376 $\pm$ 0.020 & $+$0.77 $\pm$ 0.03 & $-$1.42 $\pm$ 0.09 \\
                                  & 10.7 & J1755$-$2232 & 0.334 $\pm$ 0.027 & $+$0.78 $\pm$ 0.04 & $-$1.10 $\pm$ 0.17 \\
                                  & 11.1 & J1807$-$2506 & 0.378 $\pm$ 0.023 & $-$0.55 $\pm$ 0.04 & $-$1.61 $\pm$ 0.19 \\
                                  & 11.1 & J1755$-$2232 & 0.332 $\pm$ 0.024 & $-$0.54 $\pm$ 0.04 & $-$1.28 $\pm$ 0.20 \\
                                  \cline{2-6}
                                  & ~6.9 & {Combined} & {\bf 0.334 $\pm$ 0.020}\tablenotemark{a}  & $-$0.53 $\pm$ 0.03 & $-$1.87 $\pm$ 0.11 \\
                                  & ~7.3 &                       &                                       & $-$0.55 $\pm$ 0.03 & $-$1.83 $\pm$ 0.11 \\
                                  & ~9.0 &                       &                                       & $-$0.38 $\pm$ 0.03 & $-$2.24 $\pm$ 0.11 \\
                                  & ~9.4 &                       &                                       & $-$0.57 $\pm$ 0.03 & $-$2.20 $\pm$ 0.11 \\
                                  & 10.3 &                       &                                       & $+$0.78 $\pm$ 0.03 & $-$1.14 $\pm$ 0.11 \\
                                  & 10.7 &                       &                                       & $+$0.78 $\pm$ 0.03 & $-$1.26 $\pm$ 0.11 \\
                                  & 11.1 &                       &                                       & $-$0.54 $\pm$ 0.03 & $-$1.43 $\pm$ 0.11 \\
\hline
G011.91$-$00.61  & 39.1 & J1808$-$1822 & 0.355 $\pm$ 0.035 & $+$1.15 $\pm$ 0.11 & $-$1.14 $\pm$ 0.25 \\
                                  & 39.1 & J1825$-$1718 & 0.254 $\pm$ 0.024 & $+$1.12  $\pm$ 0.09 & $-$0.82 $\pm$ 0.49 \\
                                  \cline{2-6}
                                  & 39.1 & {Combined}    &  {\bf 0.297 $\pm$ 0.031} & $+$1.21 $\pm$ 0.10 & $-$0.99 $\pm$ 0.23 \\
\hline
G013.87$+$00.28 & $-$18.4 & J1809$-$1520 & 0.253 $\pm$ 0.024 & $+$0.12 $\pm$ 0.06 & $-$2.54 $\pm$ 0.19 \\
                                  & $-$17.9 & J1809$-$1520 & 0.224 $\pm$ 0.016 & $+$0.10 $\pm$ 0.04 & $-$2.38 $\pm$ 0.20 \\
                                  & $-$17.5 & J1809$-$1520 & 0.248 $\pm$ 0.031 & $+$0.05 $\pm$ 0.08 & $-$2.28 $\pm$ 0.23 \\
                                  & $-$16.7 & J1809$-$1520 & 0.255 $\pm$ 0.029 & $-$0.70 $\pm$ 0.07 & $-$2.64 $\pm$ 0.22 \\
                                  & $-$15.8 & J1809$-$1520 & 0.269 $\pm$ 0.019 & $-$0.62 $\pm$ 0.05 & $-$2.58 $\pm$ 0.12 \\
                                  & $-$15.0 & J1809$-$1520 & 0.274 $\pm$ 0.024 & $-$0.42 $\pm$ 0.06 & $-$2.54 $\pm$ 0.11 \\
                                  \cline{2-6}
                                  & $-$18.4 & {Combined} & {\bf 0.254 $\pm$ 0.024}\tablenotemark{a}  & $+$0.12 $\pm$ 0.06 & $-$2.54 $\pm$ 0.17 \\
                                  & $-$17.9 &                        &                                              & $+$0.10 $\pm$ 0.06 & $-$2.38 $\pm$ 0.17 \\
                                  & $-$17.5 &                        &                                              & $+$0.05 $\pm$ 0.06 & $-$2.28 $\pm$ 0.17 \\
                                  & $-$16.7 &                        &                                              & $-$0.70 $\pm$ 0.06 & $-$2.64 $\pm$ 0.17 \\
                                  & $-$15.8 &                        &                                              & $-$0.62 $\pm$ 0.06 & $-$2.59 $\pm$ 0.17 \\
                                  & $-$15.0 &                        &                                              & $-$0.42 $\pm$ 0.06 & $-$2.54 $\pm$ 0.17 \\
\hline
G016.58$-$00.05 & 62.6 & J1825$-$1551 & 0.266 $\pm$ 0.027 & $-$1.23 $\pm$ 0.04 & $-$2.35 $\pm$ 0.10 \\
                                 & 62.6 & J1825$-$1718 & 0.205 $\pm$ 0.061 & $-$1.14 $\pm$ 0.08 & $-$2.10 $\pm$ 0.20 \\
                                 & 62.6 & J1809$-$1520 & 0.219 $\pm$ 0.071 & $-$1.18 $\pm$ 0.09 & $-$2.25 $\pm$ 0.21 \\
                                 & 63.0 & J1825$-$1551 & 0.309 $\pm$ 0.038 & $-$1.17 $\pm$ 0.05 & $-$2.18 $\pm$ 0.14 \\
                                 & 63.0 & J1825$-$1718 & 0.256 $\pm$ 0.046 & $-$1.09 $\pm$ 0.06 & $-$1.89 $\pm$ 0.13 \\
                                 & 63.0 & J1809$-$1520 & 0.248 $\pm$ 0.075 & $-$1.16 $\pm$ 0.09 & $-$2.15 $\pm$ 0.21 \\
                                 & 63.4 & J1825$-$1551 & 0.306 $\pm$ 0.020 & $-$1.40 $\pm$ 0.03 & $-$3.00 $\pm$ 0.11 \\
                                 & 63.4 & J1825$-$1718 & 0.282 $\pm$ 0.061 & $-$1.33 $\pm$ 0.07 & $-$2.78 $\pm$ 0.17 \\
                                 & 63.4 & J1809$-$1520 & 0.252 $\pm$ 0.061 & $-$1.37 $\pm$ 0.08 & $-$2.96 $\pm$ 0.19 \\
                                 & 63.8 & J1825$-$1551 & 0.293 $\pm$ 0.032 & $-$1.45 $\pm$ 0.04 & $-$2.94 $\pm$ 0.10 \\
                                 & 63.8 & J1825$-$1718 & 0.226 $\pm$ 0.062 & $-$1.38 $\pm$ 0.08 & $-$2.72 $\pm$ 0.17 \\
                                 & 63.8 & J1809$-$1520 & 0.237 $\pm$ 0.056 & $-$1.42 $\pm$ 0.07 & $-$2.91 $\pm$ 0.20 \\
                                 & 64.3 & J1825$-$1551 & 0.305 $\pm$ 0.045 & $-$1.35 $\pm$ 0.06 & $-$3.00 $\pm$ 0.11 \\
                                 & 64.3 & J1825$-$1718 & 0.236 $\pm$ 0.070 & $-$1.28 $\pm$ 0.09 & $-$2.78 $\pm$ 0.18 \\
                                 & 64.3 & J1809$-$1520 & 0.246 $\pm$ 0.059 & $-$1.32 $\pm$ 0.07 & $-$2.96 $\pm$ 0.21 \\
                                  \cline{2-6}
                                 & 62.6 & {Combined} & {\bf 0.279 $\pm$ 0.023}\tablenotemark{a}  & $-$1.21 $\pm$ 0.03 & $-$2.30 $\pm$ 0.08 \\
                                 & 63.0 &                       &                                              & $-$1.14 $\pm$ 0.03 & $-$2.04 $\pm$ 0.08 \\
                                 & 63.4 &                       &                                              & $-$1.39 $\pm$ 0.02 & $-$2.94 $\pm$ 0.08 \\
                                 & 63.8 &                       &                                              & $-$1.43 $\pm$ 0.03 & $-$2.89 $\pm$ 0.07 \\
                                 & 64.3 &                       &                                              & $-$1.33 $\pm$ 0.04 & $-$2.95 $\pm$ 0.08 \\
\hline
G025.70$+$00.04 & 90.0 & J1837$-$0653 & 0.129 $\pm$ 0.020 & $-$2.91 $\pm$ 0.04 &  $-$6.25 $\pm$ 0.25 \\ 
                                  & 94.6 & J1837$-$0653 & 0.097 $\pm$ 0.021 & $-$2.87 $\pm$ 0.04 &  $-$6.05 $\pm$ 0.74 \\ 
                                  & 95.4 & J1837$-$0653 & 0.070 $\pm$ 0.046 & $-$2.87 $\pm$ 0.08 &  $-$6.32 $\pm$ 0.15 \\  
                                  \cline{2-6} 
                                  & 90.0 & {Combined} & {\bf 0.098 $\pm$ 0.029}  & $-$2.89 $\pm$ 0.05 &  $-$6.25 $\pm$ 0.35 \\ 
                                  & 94.6 &                        &                                             & $-$2.87 $\pm$ 0.05 &  $-$6.06 $\pm$ 0.35 \\  
                                  & 95.4 &                        &                                             & $-$2.90 $\pm$ 0.05 &  $-$6.32 $\pm$ 0.35 \\  

\hline
G032.04$+$00.05 & 98.0  & J1857$-$0048 & 0.204 $\pm$ 0.012\tablenotemark{b}  & $-$2.16 $\pm$ 0.03\tablenotemark{b}  & $-$4.77 $\pm$ 0.11\tablenotemark{b}  \\
                                  & 98.0  & J1904$+$0110 & 0.161 $\pm$ 0.061 & $-$2.29 $\pm$ 0.17 & $-$ 4.74 $\pm$ 0.44 \\
                                  & 98.0  & J1833$-$0323 & 0.184 $\pm$ 0.017\tablenotemark{b} & $-$2.05 $\pm$ 0.47\tablenotemark{b} & $-$5.00 $\pm$ 0.61\tablenotemark{b} \\
                                  & 98.0  & J1907$+$0127 & 0.143 $\pm$ 0.065 & $-$2.24 $\pm$ 0.18 & $-$4.73 $\pm$ 0.33 \\
                                  & 98.8  & J1857$-$0048 & 0.205 $\pm$ 0.012\tablenotemark{b}  & $-$2.17 $\pm$ 0.03\tablenotemark{b}  & $-$4.78 $\pm$ 0.11\tablenotemark{b}  \\
                                  & 98.8  & J1904$+$0110 & 0.163 $\pm$ 0.062 & $-$2.30 $\pm$ 0.17 & $-$4.75 $\pm$ 0.40 \\
                                  & 98.8  & J1833$-$0323 & 0.186 $\pm$ 0.017\tablenotemark{b} & $-$2.06 $\pm$ 0.05\tablenotemark{b} & $-$5.01 $\pm$ 0.61\tablenotemark{b} \\
                                  & 98.8  & J1907$+$0127 & 0.145 $\pm$ 0.064 & $-$2.25 $\pm$ 0.18 & $-$4.74 $\pm$ 0.30 \\
                                  & 99.5  & J1857$-$0048 & 0.208 $\pm$ 0.024 & $-$2.13 $\pm$ 0.07 & $-$4.71 $\pm$ 0.28 \\
                                  & 99.5  & J1904$+$0110 & 0.165 $\pm$ 0.038\tablenotemark{b} & $-$2.26 $\pm$ 0.10\tablenotemark{b} & $-$4.68 $\pm$ 0.30\tablenotemark{b} \\
                                  & 99.5  & J1833$-$0323 & 0.189 $\pm$ 0.042 & $-$2.03 $\pm$ 0.11 & $-$4.94 $\pm$ 0.98 \\
                                  & 99.5  & J1907$+$0127 & 0.151 $\pm$ 0.103 & $-$2.21 $\pm$ 0.28 & $-$4.67 $\pm$ 0.17 \\
                                  & 100.3  & J1857$-$0048 & 0.218 $\pm$ 0.012\tablenotemark{b}  & $-$2.09 $\pm$ 0.03\tablenotemark{b} & $-$4.74 $\pm$ 0.11\tablenotemark{b} \\
                                  & 100.3  & J1904$+$0110 & 0.175 $\pm$ 0.048 & $-$2.23 $\pm$ 0.13 & $-$4.71$\pm$ 0.43 \\
                                  & 100.3  & J1833$-$0323 & 0.199 $\pm$ 0.017\tablenotemark{b}  & $-$ 1.99 $\pm$ 0.05\tablenotemark{b}  & $-$4.97 $\pm$ 0.61\tablenotemark{b}  \\
                                  & 100.3  & J1907$+$0127 & 0.159 $\pm$ 0.078 & $-$2.17 $\pm$ 0.21 & $-$4.70 $\pm$ 0.33 \\
                                  & 101.1  & J1857$-$0048 & 0.208 $\pm$ 0.012\tablenotemark{b}  & $-$2.26 $\pm$ 0.03\tablenotemark{b} & $-$4.85 $\pm$ 0.11\tablenotemark{b} \\
                                  & 101.1  & J1904$+$0110 & 0.165 $\pm$ 0.048 & $-$2.39 $\pm$ 0.13 & $-$4.83 $\pm$ 0.43 \\
                                  & 101.1  & J1833$-$0323 & 0.188 $\pm$ 0.017\tablenotemark{b} & $-$2.16 $\pm$ 0.05\tablenotemark{b} & $-$5.08  $\pm$ 0.61\tablenotemark{b} \\
                                  & 101.1  & J1907$+$0127 & 0.148 $\pm$ 0.079 & $-$2.34 $\pm$ 0.21 & $-$4.81 $\pm$ 0.33 \\
                                  \cline{2-6}  
                                   & 98.0   & J1857$-$0048 & { 0.209 $\pm$ 0.011\tablenotemark{a}} & $-$2.16 $\pm$ 0.03 & $-$4.77 $\pm$ 0.11 \\
                                  & 98.8    &                        &                                            & $-$2.17 $\pm$ 0.03 & $-$4.78 $\pm$ 0.11 \\
                                  & 99.5    &                        &                                            & $-$2.13 $\pm$ 0.03 & $-$4.71 $\pm$ 0.11 \\
                                  & 100.3  &                        &                                            & $-$2.09 $\pm$ 0.03 & $-$4.74 $\pm$ 0.11 \\
                                  & 101.1  &                        &                                            & $-$2.26 $\pm$ 0.03 & $-$4.85 $\pm$ 0.11  \\
                                  \cline{2-6}  
                                   & 98.0   & J1904$+$0110 & { 0.166 $\pm$ 0.038\tablenotemark{a}} & $-$2.29 $\pm$ 0.10 & $-$4.74 $\pm$ 0.30 \\
                                  & 98.8    &                        &                                            & $-$2.30 $\pm$ 0.10 & $-$4.75 $\pm$ 0.30 \\
                                  & 99.5    &                        &                                            & $-$2.26 $\pm$ 0.10 & $-$4.68 $\pm$ 0.11 \\
                                  & 100.3  &                        &                                            & $-$2.23 $\pm$ 0.10 & $-$4.71 $\pm$ 0.11 \\
                                  & 101.1  &                        &                                            & $-$2.39 $\pm$ 0.10 & $-$4.82 $\pm$ 0.11  \\
                                  \cline{2-6}  
                                  & 98.0  & J1833$-$0323 & { 0.189 $\pm$ 0.016\tablenotemark{a}} & $-$2.05 $\pm$ 0.04 & $-$5.00 $\pm$ 0.61 \\
                                  & 98.8    &                        &                                            & $-$2.06 $\pm$ 0.04 & $-$5.01 $\pm$ 0.61 \\
                                  & 99.5    &                        &                                            & $-$2.03 $\pm$ 0.04 & $-$4.94 $\pm$ 0.61 \\
                                  & 100.3  &                        &                                            & $-$1.99 $\pm$ 0.04 & $-$4.97 $\pm$ 0.61 \\
                                  & 101.1  &                        &                                            & $-$2.16 $\pm$ 0.04 & $-$5.08 $\pm$ 0.61  \\
                                  \cline{2-6}  
                                                  & & Combined & {\bf 0.193 $\pm$ 0.008} & \\
\enddata 
\tablecomments{a) Formal error multiplied by $\sqrt{N_{\rm spot}}$, where $N_{\rm spot}=5$; b) Error floors decided from multi-channel (one quasar) fit (because the formal error of individual channel gave underestimated error)}
\end{deluxetable}

\clearpage

\begin{figure}
	\centering
   \includegraphics[width=7cm, angle=270]{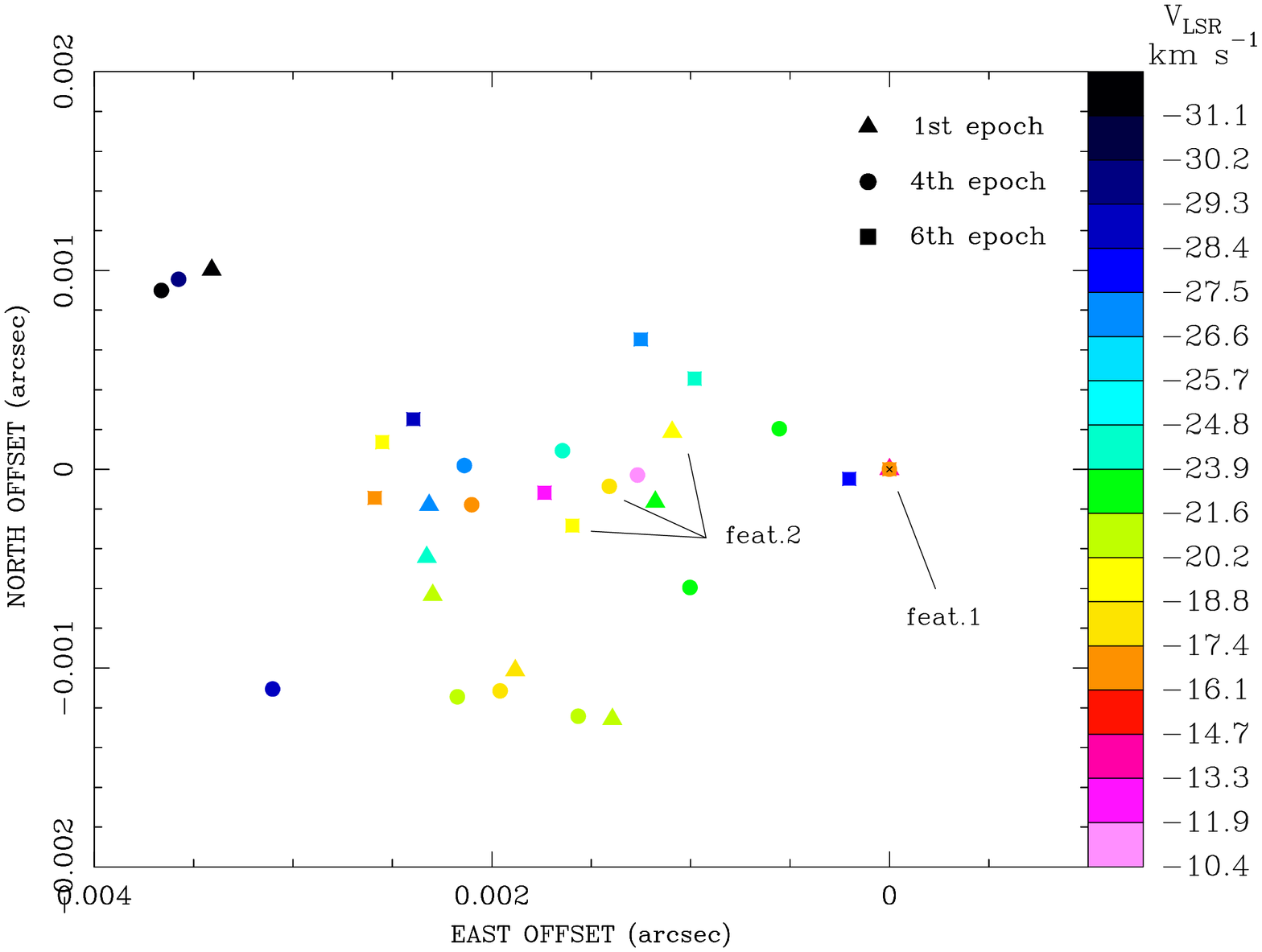}\label{fig:G013spotmap}
		\caption{\footnotesize Maser spot map for G013.87$+$00.28.}
\end{figure}

\begin{thebibliography}{}

\bibitem[Athanassoula (1992)]{Athanassoula:92}
 Athanassoula, E. 1992, \mnras, 259, 328
\bibitem[Acord, Churchwell \& Wood (1998)]{Acord:98}
 Acord, J. M., Churchwell, E., \& Wood, D. O. S. 1998, \apj, 495, 107
\bibitem[Beuther \etal (2006)]{Beuther:06}
 Beuther, H., Zhang, Q., Sridharan, T. K., Lee, C.-F., \& Zapata, L. A.
 2006, \aap, 454, 221
\bibitem[Benjamin (2009)]{Benjamin:09}
 Benjamin, R. A. 2009, in IAU Symp. 254, The Galaxy Disk in Cosmological Context,
 ed. J. Andersen, J. Bland-Hawthorn, \& B. Nordstr\"{o}m (Cambridge: Cambridge Univ. Press), 319 
\bibitem[Benjamin (2005)]{Benjamin:05}
 Benjamin, R. A., Churchwell, E., Babler, B. L., \etal 2005, \apjl, 630, L149
\bibitem[B{\l}aszkiewicz \& Kus (2004)]{Blaszkiewicz:04}
 B{\l}aszkiewicz, L., \& Kus, A. J., 2004, \aap, 413, 233
\bibitem[Bovy \etal (2012)]{Bovy:12}
 Bovy, J., \etal 2012, \apj, 759, 131
\bibitem[Brunthaler \etal (2011)]{Brunthaler:11}
 Brunthaler, A., Reid, M. J., Menten, K. M., \etal 2011, Astron.\ Nachr., 332, 461
\bibitem[Brunthaler \etal (2009)]{Brunthaler:09}
 Brunthaler, A., Reid, M. J., Menten, K. M., Zheng, X. W., Moscadelli, L., \& Xu, Y. 2009, \apj, 693, 424
\bibitem[Churchwell \etal (2009)]{Churchwell:09}
Churchwell, E., Babler, B. L., Meade, M. R., \etal 2009, \pasp, 121, 213
\bibitem[Clemens \etal (1988)]{Clemens:88}
 Clemens, D. P., Sanders, D. B., \& Scoville, N. Z. 1988, \apj, 327, 139
\bibitem[Cohen \etal (1980)]{Cohen:80}
 Cohen, R. S., Cong, H., Dame, T. M., \& Thaddeus, P. 1980 \apjl, 239, L53
\bibitem[Contopoulos \& Papayannopoulos (1980)]{Contopoulos:80}
 Contopoulos, G., \& Papayannopoulos, T. 1980, \aap, 92, 33
\bibitem[Cyganowski \etal (2009)]{Cyganowski:09}
 Cyganowski, C. J., Brogan, C. L., Hunter, T. R., \&Churchwell, E.
 2009, \apj, 702, 1615
\bibitem[Cyganowski \etal (2011)]{Cyganowski:11}
 Cyganowski, C. J., Brogan, C. L., Hunter, T. R., Churchwell, E., \& Zhang, Q.
 2011,  \apj, 729, 124
\bibitem[Cyganowski \etal (2008)]{Cyganowski:08}
 Cyganowski, C. J., Whitney, B. A., Holden, E., \etal 2008, \aj, 136, 2391
\bibitem[Dame \etal (1986)]{Dame:86}
 Dame, T. M., Elmegreen, B. G., Cohen, R. S., \& Thaddeus, P. 1986, \apj, 305, 892
\bibitem[Dame \etal (2001)]{Dame:01}
 Dame, T. M., Hartmann, D., \& Thaddeus, P.  2001, \apj, 547, 792 
\bibitem[Dame \& Thaddeus (2011)]{Dame:11}
 Dame, T. M., \& Thaddeus, P.  2011, \apjl, L734
\bibitem[Davis \etal (2012)]{Davis:12}
 Davis, B. L., Berrier, J. C., Shields, D. W., Kennefick, J.,
 Kennefick, D., Seigar, M. S., Lacy, C. H. S., \& Puerari, I.
 2012 \apjs, 199, 33 
\bibitem[Deller \etal (2007)]{Deller:07}
 Deller, A. T., Tingay, S. J., Bailes, M., \& West, C.
 2007, \pasp, 119, 318
\bibitem[Drimmel (2000)]{Drimmel:00}
 Drimmel, R. 2000, \aap, 358, L13
\bibitem[Drimmel \& Spergel (2001)]{Drimmel:01}
 Drimmel, R., \& Spergel, D. N. 2001, \apj, 556, 181
\bibitem[Ellingsen \etal (2011)]{Ellingsen:11}
 Ellingsen, S. P., Breen, S. L., Sobolev, A. M., Voronkov, M. A., Caswell, J. L., \& Lo, N.
 2011, \apj, 742, 109
\bibitem[Ellsworth-Bowers \etal (2013)]{Ellsworth-Bowers:13}
 {{Ellsworth-Bowers}, T.~P., {Glenn}, J., {Rosolowsky}, E., {Mairs}, S., {Evans}, II, N.~J.,
  {Battersby}, C., {Ginsburg}, A., {Shirley}, Y.~L., \& {Bally}, J.}  2013, \apj, 770, 39
\bibitem[Feldt \etal (2003)]{Feldt:03}
 Feldt, M., Puga, E., Lenzen, R., Henning, T., Brandner, W., Stecklum, B.,
 Lagrange, A.-M., Gendron, E., \& Rousset, G.  2003, \apj, 599, L91
\bibitem[Fey \etal (2004)]{Fey:04}
  Fey, A. L., Ma, C., Arias, E. F., Charlot, P.,
  Feissel-Vernier, M., Gontier, A.-M., Jacobs, C. S., Li, J.,
  \& MacMillan, D. S.  2004, \aj, 127, 3587
\bibitem[Fish \etal (2003)]{Fish:03}
 Fish, V. L., Reid, M. J., Wilner, D. J., \& Churchwell, E.
 2003, \apj, 587, 701 
\bibitem[Gaylard \etal (1994)]{Gaylard:94}
 Gaylard, M. J., MacLeod, G. C., \& van der Walt, D. J.
 1994, \mnras, 269, 257
\bibitem[Gerhard (2002)]{Gerhard:02}
 Gerhard, O. 2002, ASP Conf.\ Ser.\ 273,
 The Dynamics, Structure \& History of Galaxies, ed.\ G. S. Da Costa, \& H. Jerjen
  (San Francisco: ASP), 73
\bibitem[Greisen (2003)]{Greisen:03}
  Greisen, E. W.  2003, in Information Handling in Astronomy:
  Historical Vistas, ed.\ A.~Heck (Dordrecht: Kluwer), 109
\bibitem[Haschick \etal (1989)]{Haschick:89}
 Haschick, A. D., Baan, W. A., Menten, K. M. 1989, \apj, 346, 330
\bibitem[Hammersley \etal (2000)]{Hammersley:00}
 Hammersley, P. L., Garz\'{o}n, F., Mahoney, T. J., L\'{o}pez-Corredoira, M., \& Torres, M. A. P.
 2000, \mnras, 317, L45
\bibitem[Hofner \& Churchwell (1996)]{Hofner:96}
  Hofner, P. \& Churchwell, E.  1996, \aaps, 120, 283
\bibitem[Honma \etal (2013)]{Honma:13}
 Honma, M., Nagayama, T., Ando, K., \etal  2013, \pasj, in press (arXiv:1211.3843)
\bibitem[Honma \& Sofue (1997)]{Honma:97}
 Honma, M., \& Sofue, Y. 1997, \pasj, 49, 453
\bibitem[Hunter \etal (2008)]{Hunter:08}
  Hunter, T. R., Brogan, C. L., Indebetouw, R., \& Cyganowski, C. J. 2008, \apj, 680, 1271
\bibitem[Immer \etal (2011)]{Immer:11}
  Immer, K., Brunthaler, A., Reid, M. J., \etal 2011, \apjs, 194, 25
\bibitem[Immer \etal (2013)]{Immer:13}
  Immer, K., Reid, M. J., Menten, K. M., Brunthaler, A., \& Dame, T. M.
  2013, \aap 553, 117
\bibitem[Jaffe \etal (1981)]{Jaffe:81}
 Jaffe, D. T., Guesten, R., \& Downes, D. 1981, \apj, 250, 621
\bibitem[Jackson \etal (2006)]{Jackson:06}
 Jackson, J. M., Rathborne, J. M., Shah, R. Y., \etal 2006, \apjs, 163, 145
\bibitem[Kettenis \etal (2006)]{Kettenis:06}
 Kettenis, M., van Langevelde, H. J., Reynolds, C., \& Cotton, B.  
 2006, in ASP Conf.\ Ser.\ 351, Astronomical Data Analysis Software and Systems XV,
 ed.\ C.\ Gabriel, C.\ Arviset, D.\ Ponz, \& S.\ Enrique (San Francisco, CA: ASP), 497
\bibitem[Ma \etal (2009)]{Ma:09}
 Ma, C., Arias, E.\ F., Bianco, G., \etal 2009, in The Second Realization of the International
 Celestial Reference Frame by Very Long Baseline Interferometry
 ed.\ A.\ L.\ Fey, D.\ Gordon, \& C.\ S.\ Jacobs
 (IERS Technical Note No.\  35; Frankfurt am Main, Germany: IERS)
\bibitem[Motogi \etal (2011)]{Motogi:11}
 Motogi, K., Sorai, K., Habe, A., Honma, M., Kobayashi, H., Sato, K. 
 2011 \pasj, 63, 31
\bibitem[Nakanishi \& Sofue (2006)]{Nakanishi:06}  
 Nakanishi, H., \& Sofue, Y. 2006, \pasj, 58, 847
\bibitem[Petrov \etal (2005)]{Petrov:05}
  Petrov, L., Kovalev, Y. Y., Fomalont, E. \& Gordon, D.
  2005, \aj, 129, 1163
\bibitem[Reid \etal (2009a)]{Reid:09I} 
 Reid, M. J., Menten, K. M., Brunthaler, A., Zheng, X. W.,
 Moscadelli, L. \& Xu, Y.  2009a, \apj, 693, 397
\bibitem[Reid \etal (2009b)]{Reid:09VI} 
 Reid, M. J. \etal  2009b, \apj, 700, 137
\bibitem[Reid \etal (2014)]{Reid:13} 
 Reid, M. J., Menten, K. M., Zheng, X. W., \etal 2014, \apj, 783, 130
\bibitem[Sanna \etal (2010)]{Sanna:10} 
 Sanna, A., Moscadelli, L., Cesaroni, R., Tarchi, A., Furuya, R. S., \& Goddi, C.
 2010, \aap, 517, 71
\bibitem[Sanna \etal (2012)]{Sanna:12} 
 Sanna, A., Reid, M. J., Dame, T. M., Menten, K. M., Brunthaler, A.,
 Moscadelli, L., Zheng, X. W., \& Xu, Y.  2012, \apj, 745, 82 
\bibitem[Sanna \etal (2009)]{Sanna:09} 
 Sanna, A., Reid, M. J., Moscadelli, L., Dame, T. M., Menten, K. M., Brunthaler, A., Zheng, X. W., \& Xu, Y.
 2009 \apj, 706, 464
\bibitem[Sato \etal (2010a)]{Sato:10G}
 Sato, M., Hirota, T., Reid, M. J., Honma, M., Kobayashi, H., 
 Iwadate, K., Miyaji, T. \& Shibata, K. M.  2010a,
 \pasj, 62, 287
\bibitem[Sato \etal (2010b)]{Sato:10W}
 Sato, M., Reid, M. J., Brunthaler, A. \& Menten, K. M.
 2010b, \apj, 720, 1055
\bibitem[Sch\"{o}nrich \etal (2010)]{Schoenrich:10} 
 Sch\"{o}nrich, R., Binney, J., \& Dehnen, W.  2010,  \mnras, 403, 1829
\bibitem[Scoville \& Solomon (1975)]{Scoville:75}
 Scoville, N. Z., \& Solomon, P. M. 1975 \apjl, 199, L105
\bibitem[Shane (1972)]{Shane:72} 
 Shane, W. W. 1972, \aap, 16, 118 
\bibitem[Sofue (2006)]{Sofue:06}
 Sofue, Y. 2006 \pasj, 58, 335 
\bibitem[Sridharan \etal (2002)]{Sridharan:02} 
 Sridharan, T. K., Beuther, H., Schilke, P., Menten, K. M., \& Wyrowski, F.
 2002, \apj, 566, 931
\bibitem[Sweilo \etal (2004)]{Sweilo:04} 
 Sewilo, M., Watson, C., Araya, E., Churchwell, E., Hofner, P., \& Kurtz, S.
 2004, \apjs, 154, 553
\bibitem[Szymczak \etal (2004)]{Szymczak:04} 
 Szymczak, M., Gerard, E. 2004, \aap, 414, 235
\bibitem[Szymczak \etal (2005)]{Szymczak:05}
 Szymczak, M., Pillai, M., \& Menten, K. M. et al. 2005, \aap, 434, 613 
\bibitem[Taylor \& Cordes (1993)]{Taylor:93}
 Taylor, J. H., \& Cordes, J. M. 1993, \apj, 411, 674 
\bibitem[Thompson \etal (2006)]{Thompson:06} 
 Thompson, M. A., Hatchell, J., Walsh, A. J., MacDonald, G. H., \& Millar, T. J.
 2006, \aap, 453, 1003
\bibitem[van Albada \& Sanders (1982)]{vanAlbada:82}
 van Albada, T. S, \& Sanders, R. H. 1982, \mnras, 201, 303
\bibitem[van der Walt (1995)]{vanderWalt:95} 
  van der Walt, D. J., Gaylard, M. J., \& MacLeod, G. C.
  1995, \aap, 110, 81
\bibitem[Vel\'{a}zquez \etal (2002)]{Velazquez:02} 
 Vel\'{a}zquez, P. F., Dubner, G. M., Goss, W. M., \& Green, A. J.
 2002, \aj, 124, 2145
\bibitem[Walsh \etal (1997)]{Walsh:97}
Walsh, A. J., Hyland, A. R., Robinson, G., \& Burton, M. G., 1997, \mnras, 291, 261 
\bibitem[Westerhout (1957)]{Westerhout:57}
 Westerhout, G. 1957, Bull.\ Astro.\ Inst.\ Netherlands, 13, 201 (No.\ 475)
\bibitem[Wood \& Churchwell (1989)]{Wood:89} 
 Wood, D. O. S., \& Churchwell, E.  1989, \apjs, 69, 831
\bibitem[Wu \etal (2014)]{Wu:13} 
 Wu, Y. W., Sato, M., Reid, M. J., \etal, 2014, \aap, 566, 17
\bibitem[Xu \etal (2009)]{Xu:09} 
  Xu, Y., Reid, M. J., Menten, K. M., Zheng, X. W.
  2006, \apjs, 166, 526
\bibitem[Xu \etal (2011)]{Xu:11} 
  Xu, Y., Moscadelli, L., Reid, M. J., Menten, K. M., Zhang, B., Zheng, X. W., \& Brunthaler, A.
  2011, \apj, 733, 25
\bibitem[Xu \etal (2013)]{Xu:13} 
 Xu, Y., Li, J. J., Reid, M. J., Menten, K. M., Zheng, X. W., Brunthaler, A., Moscadelli, L., Dame, T. M., \& Zhang, B.
 2013, \apj, 769, 15
\bibitem[Zhang \etal (2013)]{Zhang:13a} 
   Zhang, B., Reid, M. J., Menten, K. M., \etal, 2013a, \apj, 775, 79
\bibitem[Zhang \etal (2014)]{Zhang:13b} 
   Zhang, B., Reid, M. J., \etal, 2014, \apj, 781, 89
  \end{thebibliography}
\end{document}